\documentclass[useAMS,usenatbib]{mn2e}
\usepackage{epsfig}

\title[Faint emission lines in M16, M20 and NGC~3603]{Faint emission lines in the Galactic H~{\sc ii} regions 
M16, M20 and NGC~3603 
\thanks{Based on observations collected at the European 
Southern
Observatory, Chile, proposal number ESO 68.C-0149(A)}}
\author[J. Garc\'\i a-Rojas et al.]
  {J.~Garc\'\i a-Rojas,$^1$\thanks{E-mail: jogarcia@iac.es}
  C.~Esteban,$^1$   M.~Peimbert,$^2$ M.T. Costado,$^3$
  \newauthor 
  M.~Rodr\'\i guez,$^4$ A.~Peimbert,$^2$ M.T.~Ruiz,$^5$ \\
  $^1$Instituto de Astrof\'\i sica de Canarias, E-38200 La Laguna, Tenerife, Spain \\
  $^2$Instituto de Astronom\'\i a, UNAM, Apdo. Postal 70-264, M\'exico 04510 D.F., Mexico\\
  $^3$Departamento de Astrof\'\i sica, Universidad de La Laguna, La Laguna, Spain\\
  $^4$Instituto Nacional de Astrof\'\i sica, \'Optica y Electr\'onica INAOE, Apdo. Postal 51 y 216, 
  7200 Puebla, Pue., Mexico\\
  $^5$Departamento de Astronom\'\i a, Universidad de Chile, Casilla Postal 36D, Santiago de Chile,
  Chile\\}

\newcommand{\elecd}{$n_{\rm e}$}
\newcommand{\te}{$T_{\rm e}$}
\newcommand{\hb}{H$\beta$}
\newcommand{\ha}{H$\alpha$}

\newcommand{\iii}{{\sc iii}}
\newcommand{\foiii}{[O~{\sc iii}]}

\newcommand{\foii}{[O~{\sc ii}]}
\newcommand{\fsii}{[S~{\sc ii}]}
\newcommand{\fsiii}{[S~{\sc iii}]}
\newcommand{\fni}{[N~{\sc i}]}
\newcommand{\fnii}{[N~{\sc ii}]}
\newcommand{\fariv}{[Ar~{\sc iv}]}
\newcommand{\fcliii}{[Cl~{\sc iii}]}
\newcommand{\fcliv}{[Cl~{\sc iv}]}
\newcommand{\fneiii}{[Ne~{\sc iii}]}
\newcommand{\ffeii}{[Fe~{\sc ii}]}
\newcommand{\ffeiii}{[Fe~{\sc iii}]}

\newcommand{\oiii}{O~{\sc iii}}
\newcommand{\nitroi}{N~{\sc i}}
\newcommand{\nii}{N~{\sc ii}}

\newcommand{\sili}{Si~{\sc i}}
\newcommand{\silii}{Si~{\sc ii}}
\newcommand{\oi}{O~{\sc i}}
\newcommand{\oii}{O~{\sc ii}}
\newcommand{\ci}{C~{\sc i}}
\newcommand{\cii}{C~{\sc ii}}

\newcommand{\sii}{S~{\sc ii}}
\newcommand{\siii}{S~{\sc iii}}

\newcommand{\cliii}{Cl~{\sc iii}}
\newcommand{\fariii}{[Ar~{\sc iii}]}
\newcommand{\fei}{Fe~{\sc i}}

\newcommand{\feiii}{Fe~{\sc iii}}

\newcommand{\ari}{Ar~{\sc i}}
\newcommand{\ariii}{Ar~{\sc iii}}

\newcommand{\hi}{H\,{\sc i}}
\newcommand{\hii}{H~{\sc ii}}
\newcommand{\di}{D\,{\sc i}}
\newcommand{\hei}{He~{\sc i}}
\newcommand{\heii}{He~{\sc ii}}

\newcommand{\ts}{\emph{$t^2$}}
\newcommand{\mc}{\multicolumn}

\begin{document}

\maketitle

\begin{abstract}

We present deep echelle spectrophotometry of the Galactic {\hii} regions M16, M20 and NGC~3603. The data have been taken 
with the Very Large Telescope Ultraviolet-Visual Echelle Spectrograph in the 3100 to 10400 \AA\ range.
We have detected more than 200 emission lines in each region.
Physical conditions have been derived using different continuum and line intensity ratios. 
We have derived He$^{+}$, C$^{++}$ and O$^{++}$ abundances from pure 
recombination lines as well as abundances from collisionally excited lines for 
a large number of ions of different elements. We have obtained consistent 
estimations of the temperature fluctuation parameter, {\ts}, using different methods. We also report the detection of deuterium Balmer lines up to D$\delta$ (M16) 
and to D$\gamma$ (M20) in the blue wings of the hydrogen lines, which excitation mechanism seems to be continuum fluorescence. 
The temperature fluctuations paradigm agree with the results obtained from optical CELs  and the more uncertain ones from far 
IR fine structure CELs in NGC~3603, although, more observations covering the same volume of the nebula are 
necessary to obtain solid conclusions.
\end{abstract}


\section{Introduction}\label{intro}

Spectrophotometric studies of Galactic {\hii} regions provide paramount information for the study of 
the chemical evolution of our galaxy. The use of echelle spectrographs at large aperture telescopes is 
a step ahead in the knowledge of the physics and chemical composition of these objects. In the last years, 
our group and others have obtained deep 
intermediate and high resolution optical spectra of bright Galactic {\hii} regions \citep[e.g.][]{peimbertetal93,
estebanetal98,estebanetal99a,estebanetal99b,estebanetal04,garciarojasetal04,garciarojasetal05,tsamisetal03}, 
and extragalactic {\hii} regions \citep[e.g.][]{estebanetal02, apeimbert03, tsamisetal03}. These observations have permitted to 
obtain accurate measurements of very faint recombination lines (hereinafter RLs) of heavy element ions 
(especially {\cii} and {\oii} lines) avoiding the problem of line blending. A common result of all the studies 
commented above is that the abundance determinations from RLs are systematically larger than 
those obtained using the standard methods based on the standard analysis of collisionally excited lines 
(hereinafter CELs). The discrepancies can be of the order of a factor 2-3. This problem may be related to 
the so-called temperature fluctuations 
suggested to be present in ionised nebulae \citep{torrespeimbertetal80}. In fact, the intensity of CELs is 
much more strongly dependent on the temperature 
than RLs. This implies that RLs should be, in principle, more precise indicators of the true chemical 
abundances of the nebula. Nonetheless, this is still a controversial matter that has not been solved; in fact, 
alternative causes have been proposed recently, such as e.g. small-scale chemical inhomogeneities \citep{tsamisetal03, 
tsamispequignot05}. 

This paper presents results of Very Large Telescope (VLT) spectrophotometry obtained with the 
Ultraviolet-Visual Echelle Spectrograph (UVES) of three bright {\hii} regions: M16, M20 and NGC~3603. 
Our dataset, of an unprecedented quality, allows us to derive, with high precision, physical conditions and 
ionic abundances of many heavy elements and, for the first time in the three {\hii} regions studied in 
this paper, C$^{++}$ and O$^{++}$ abundances from pure recombination lines.

M16 --{\it the Aquila nebula}, NGC~6611--  is a well known {\hii} region of our galaxy. Several works about 
star formation in the ``elephant trunks'' have appeared in the last years, most of them related to the evaporating gaseous globules (EGGs) discovered in this object \citep{hesteretal96,thompsonetal02,mccaughreanandersen02}
In particular, recent near-infrared observations have led some authors to suggest that the epoch of star 
formation in M16 may be near 
its endpoint \citep{thompsonetal02,mccaughreanandersen02}.
Surprisingly, there are only a few optical spectrophotometric works on the chemical composition of this object, 
computing the most common abundance ratios \citep{hawley78, rodriguez99b,deharvengetal00}.  

M20 --{\it the Trifid nebula}, NGC~6514-- is a nearby, small and symmetrical {\hii} region ionised by the 
O7 V star HD 164492A. Although many works have been carried out to understand its kinematics 
\citep[e.g.][]{bohuski73a,rosadoetal99} and its temperature and density structure \citep[e.g.][]{bohuski73b,dopita74,copettietal00}, 
there is only a handful of studies devoted to the chemical composition of this nebula. \citet{hawley78} 
and \citet{rodriguez99b} derived heavy element 
abundances for M20 from spectrophotometric data. \citet{lyndsoneil85} used long-slit spectroscopic 
data and narrow-band photometry to measure the intensities of several bright emission lines 
across the nebula, and derived He, N, O and S abundances for M20 through the computing of a dusty model. 
Additionally, because of the particular shape, size and dust distribution of this nebula some works were carried out 
to explore the interaction between gas and dust in this region \citep[e.g.][]{krishnaswamyodell67}

Finally, NGC~3603 is the only optically visible, giant {\hii} region in our Galaxy \citep{gossradhakrishnan69,balicketal80}. 
The study of the physical properties of this object 
is crucial for the knowledge of physical processes in large {\hii} regions. 
Many efforts have been developed on the study of the star formation and the stellar content of 
NGC~3603 \citep[e.g.][]{brandletal99,stolteetal04} and 
on the study of its kinematics \citep[e.g.][]{balicketal80,clayton86, clayton90, nurnbergeretal02}, but few works have studied 
the chemical properties of the ionised gas of this object using optical data \citep{melnicketal89,girardietal97,tapiaetal01}
and mid and far-infrared data \citep{lacyetal82, simpsonetal95}.

In \S\S~\ref{obsred} and~\ref{lin} we describe the observations, the data reduction procedure and 
the measurement and identification of the emission lines. 
In \S~\ref{phiscond} we calculate electron temperatures 
and densities using several diagnostic ratios and discuss the {\ts} results. 
In \S~\ref{abundances} ionic abundances are determined based on both kinds of lines: CELs and RLs. 
Total abundances are determined in \S~\ref{totabun}. Deuterium Balmer lines and their excitation mechanism are discussed in 
\S~\ref{deut}. The detection of velocity components in NGC~3603 is reported in \S~\ref{veloc}.
Finally, in \S\S~\ref{discus} and ~\ref{conclu} we present the general discussion and the conclusions, 
respectively.

\section{Observations and Data Reduction}\label{obsred}

\setcounter{table}{0}
\begin{table}
\begin{minipage}{75mm}
\centering \caption{Journal of observations.}
\label{tobs}
\begin{tabular}{l@{\hspace{2.8mm}}c@{\hspace{2.8mm}}c@{\hspace{2.8mm}}c@{\hspace{2.8mm}}c@{\hspace{2.8mm}}}
\noalign{\hrule} \noalign{\vskip3pt}
Date & $\Delta\lambda$~(\AA) & \mc{3}{c}{Exp. time (s)}\\
& & M20 & M16 & NGC~3603 \\
\noalign{\vskip3pt} \noalign{\hrule} \noalign{\vskip3pt}
2003/03/29-31 & 3000$-$3900 & 3$\times$600 & 3$\times$900 & 3$\times$600 \\
& 3800$-$5000 & 3$\times$1800 & 3$\times$1825 & 60, 3$\times$1320 \\
& 4750$-$6800 & 3$\times$600 & 3$\times$900 & 3$\times$600 \\
& 6700$-$10400 & 3$\times$1800 & 3$\times$1825 & 60, 3$\times$1320 \\
\noalign{\vskip3pt} \noalign{\hrule} \noalign{\vskip3pt}
\end{tabular}
\end{minipage}
\end{table}

The observations were made on 2003 March 29, 30 and 31 with UVES 
\citep{dodoricoetal00}, at the VLT Kueyen Telescope in Cerro Paranal Observatory (Chile). We used 
the standard settings in both the red and 
blue arms of the spectrograph, covering the spectral region from 3100 to 10400 \AA. The log of the 
observations is presented in Table~\ref{tobs}.

The wavelength regions 5783--5830 \AA\ and 8540--8650
\AA\ were not observed due to a gap between the two CCDs used in
the red arm. There are also five small gaps that were not observed, 9608--9612 \AA, 9761--9767 \AA, 
9918--9927 \AA, 10080--10093 \AA\ and 10249--10264 \AA, because the five redmost orders did not fit completely within the CCD.  

The slit was oriented east-west in all the cases and the atmospheric dispersion corrector (ADC) was 
used to keep the same observed
region within the slit regardless of the air mass value.  The slit width was
set to 3$\arcsec$ and the slit length was set to 10$\arcsec$ in the blue arm and to 12$\arcsec$ 
in the red arm; the slit width was chosen to maximize the S/N ratio of the
emission lines and to maintain the required resolution to separate most of the
weak lines needed for this project. The effective resolution
at a given wavelength is approximately $\Delta \lambda \sim \lambda / 8800$. 

The centre of the slit was placed 48$\arcsec$ north and 40$\arcsec$ west of BD-13 4930, for M16; 
17$\arcsec$ north and 10$\arcsec$ east of HD164492, for M20; and 12$\arcsec$ north and 116$\arcsec$ east 
of HD 306201, for NGC~3603. All slit positions cover very bright zones of the nebulae. The final 
usable one-dimensional spectra were extracted from an area 
of 3$\arcsec$$\times$8.5$\arcsec$ for all the objects.
  
The spectra were reduced using the {\sc IRAF}\footnote{{\sc IRAF} is distributed by NOAO, which is operated by AURA,
under cooperative agreement with NSF.} echelle reduction
package, following the standard procedure of bias subtraction, aperture
extraction, flatfielding, wavelength calibration and flux calibration. 
The standard stars EG 247, C-32d9927 \citep{hamuyetal92,hamuyetal94} and HD 49798 \citep{turnsheketal90,bohlinlindler92} were observed for flux calibration.

\section{Line Intensities and Reddening Correction}\label{lin}

Line intensities were measured integrating all the flux in the line between two 
given limits and over a local continuum estimated by eye. In the cases of line blending, a multiple 
Gaussian profile fit procedure was applied to obtain the line flux of each 
individual line. These measurements were made with the {\sc SPLOT} routine of the {\sc 
IRAF} 
package. In some cases of very tight blends or blends with very bright telluric lines the 
analysis was performed via Gaussian fitting (or Voigt profiles in the case of sky emission lines) 
making use of the Starlink {\sc DIPSO} software 
\citep{howardmurray90}. Also, {\sc DIPSO} was used to obtain the best values for the fluxes 
of most of the {\hi} Balmer lines in M20, which were affected by absorptions associated to the 
stellar continuum scattered light. For each single or multiple Gaussian fit, {\sc DIPSO} 
gives the fit parameter (radial velocity centroid, Gaussian sigma, {\sc FWHM}, etc.) 
and their associated statistical errors.

\setcounter{table}{1}
\begin{table}
\begin{minipage}{75mm}
\centering \caption{Observed and reddening-corrected relative line fluxes $F(\lambda)$ and $I(\lambda)$ 
respectively, in units such that F(H$\beta$)=100.}
\label{lineid}
\begin{tabular}{l@{\hspace{2.8mm}}l@{\hspace{2.8mm}}c@{\hspace{1.8mm}}
c@{\hspace{2.8mm}}c@{\hspace{2.8mm}}c@{\hspace{2.8mm}}r@{\hspace{1.8mm}}l@{\hspace{1.8mm}}}
\noalign{\hrule} \noalign{\vskip3pt}
$\lambda_{\rm 0}$& & & $\lambda_{\rm obs}$& & & err & Notes \\
(${\rm \AA}$)& Ion& Mult.& (${\rm \AA}$)& $F(\lambda)$& $I(\lambda)$$^{\rm a}$& (\%)$^{\rm b}$ & \\
\noalign{\vskip3pt} \noalign{\hrule} \noalign{\vskip3pt}
\multicolumn{8}{c}{\bf \large{M 16}}\\
\noalign{\vskip3pt} \noalign{\hrule} \noalign{\vskip3pt}
3187.84     &  He I       &  3	    & 3187.81  &    0.693  &	1.352  &   6  &  \\
3354.42     &  He I       &  8	    & 3354.64  &    0.146  &	0.276  &  18  &  \\
3347.59     &  He I       &  7	    & 3447.65  &    0.201  &	0.375  &  14  &  \\
3554.42     &  He I       &  34     & 3554.45  &    0.219  &	0.401  &  13  &  \\
3587.28     &  He I       &  32     & 3587.37  &    0.178  &	0.325  &  15  &  \\
\end{tabular}
\end{minipage}
\end{table}

\setcounter{table}{1}
\begin{table}
\begin{minipage}{75mm}
\centering \caption{continued}
\begin{tabular}{l@{\hspace{2.8mm}}l@{\hspace{2.8mm}}c@{\hspace{1.8mm}}
c@{\hspace{2.8mm}}c@{\hspace{2.8mm}}c@{\hspace{2.8mm}}r@{\hspace{1.8mm}}l@{\hspace{1.8mm}}}
\noalign{\hrule} \noalign{\vskip3pt}
$\lambda_{\rm 0}$& & & $\lambda_{\rm obs}$& & & err & Notes \\
(${\rm \AA}$)& Ion& Mult.& (${\rm \AA}$)& $F(\lambda)$& $I(\lambda)$$^{\rm a}$& (\%)$^{\rm b}$ & \\
\noalign{\vskip3pt} \noalign{\hrule} \noalign{\vskip3pt}
4120.82     &  He I       &  16     & 4120.99  &    0.125  &	0.186  &  10  &  \\
4143.76     &  He I       &  53     & 4143.86  &    0.217  &	0.318  &   6  &  \\
4267.15     &  C II       &  6	    & 4267.23  &    0.198  &	0.272  &   7  &  \\
4303.61     &  O II       &  65a    & 4303.88  &    0.043  &	0.058  &  23  &  \\
4303.82     &  O II       &  53a    &	       & 	   &	       &      &  \\
4319.63     &  O II       &  2	    & 4319.75  &    0.012  &	0.016  &  :   &  \\
\end{tabular}
\end{minipage}
\end{table}

\setcounter{table}{1}
\begin{table}
\begin{minipage}{75mm}
\centering \caption{continued}
\begin{tabular}{l@{\hspace{2.8mm}}l@{\hspace{2.8mm}}c@{\hspace{1.8mm}}
c@{\hspace{2.8mm}}c@{\hspace{2.8mm}}c@{\hspace{2.8mm}}r@{\hspace{1.8mm}}l@{\hspace{1.8mm}}}
\noalign{\hrule} \noalign{\vskip3pt}
$\lambda_{\rm 0}$& & & $\lambda_{\rm obs}$& & & err & Notes \\
(${\rm \AA}$)& Ion& Mult.& (${\rm \AA}$)& $F(\lambda)$& $I(\lambda)$$^{\rm a}$& (\%)$^{\rm b}$ & \\
\noalign{\vskip3pt} \noalign{\hrule} \noalign{\vskip3pt}
5686.21     &  N II       &  3	    & 5686.18  &    0.041  &	0.028  &  :  &  c \\
5710.76     &  N II       &  3	    & 5710.88  &    0.022  &	0.015  &  :  &  \\
5754.64     &  [N II]     &  3F     & 5754.78  &    2.110  &	1.403  &   4  &  \\
5875.64     &  He I       &  11     & 5875.78  &   18.461  &   11.636  &   3  &  \\
5978.83	    &  Si II      &  4      & 5979.09  &    0.078  &    0.047  &   :  &  \\
6046.44     &  O I        &  22     & 6046.71  &    0.052  &	0.030  &  :  &  \\
\end{tabular}
\end{minipage}
\end{table}

\setcounter{table}{1}
\begin{table}
\begin{minipage}{75mm}
\centering \caption{continued}
\begin{tabular}{l@{\hspace{2.8mm}}l@{\hspace{2.8mm}}c@{\hspace{1.8mm}}
c@{\hspace{2.8mm}}c@{\hspace{2.8mm}}c@{\hspace{2.8mm}}r@{\hspace{1.8mm}}l@{\hspace{1.8mm}}}
\noalign{\hrule} \noalign{\vskip3pt}
$\lambda_{\rm 0}$& & & $\lambda_{\rm obs}$& & & err & Notes \\
(${\rm \AA}$)& Ion& Mult.& (${\rm \AA}$)& $F(\lambda)$& $I(\lambda)$$^{\rm a}$& (\%)$^{\rm b}$ & \\
\noalign{\vskip3pt} \noalign{\hrule} \noalign{\vskip3pt}
8247.73     &  H I        &  P41    & 8248.00  &    0.168  &	0.043  &  10  & c \\
8249.97     &  H I        &  P40    & 8250.20  &    0.134  &	0.034  &  11  &  \\
8252.40     &  H I        &  P39    & 8252.62  &    0.175  &	0.044  &  10  &  \\
8255.02     &  H I        &  P38    & 8255.28  &    0.207  &	0.052  &   9  &  \\
8257.85     &  H I        &  P37    & 8258.12  &    0.161  &	0.041  &  10  &  \\
8260.93     &  H I        &  P36    & 8261.21  &    0.201  &	0.051  &   9  &  \\
\end{tabular}
\end{minipage}
\end{table}

\setcounter{table}{1}
\begin{table}
\begin{minipage}{75mm}
\centering \caption{continued}
\begin{tabular}{l@{\hspace{2.8mm}}l@{\hspace{2.8mm}}c@{\hspace{1.8mm}}
c@{\hspace{2.8mm}}c@{\hspace{2.8mm}}c@{\hspace{2.8mm}}r@{\hspace{1.8mm}}l@{\hspace{1.8mm}}}
\noalign{\vskip3pt} \noalign{\hrule} \noalign{\vskip3pt}
$\lambda_{\rm 0}$& & & $\lambda_{\rm obs}$& & & err & Notes \\
(${\rm \AA}$)& Ion& Mult.& (${\rm \AA}$)& $F(\lambda)$& $I(\lambda)$$^{\rm a}$& (\%)$^{\rm b}$ & \\
\noalign{\vskip3pt} \noalign{\hrule} \noalign{\vskip3pt}
9210.28     &  He I       &  6/9    & 9210.55  &    0.347  &	0.068  &   9  &  \\
9229.01     &  H I        &  P9     & 9229.26  &   11.076  &	2.161  &   8  &  \\
9405.73     &  C I        & $^1$P$^0$--$^1$D & 9406.11  &    0.127  &	0.024  &  12  & g \\
9463.57     &  He I       &  1/5    & 9463.84  &    0.302  &	0.056  &   9  &  \\
9516.57     &  He I       &  4/7    & 9516.70  &    0.262  &	0.048  &  10  &  \\
9526.16     &  He I       &  6/8    & 9526.55  &    0.386  &	0.071  &   9  &  \\
9530.60     &  [S III]    &  1F     & 9531.21  &  321.963  &   59.257  &   8  &  \\
9545.97     &  H I        &  P8     & 9546.21  &   13.877  &	2.547  &   8  & d \\
9702.62     &  He I       &  75     & 9703.34  &    0.086  &	0.015  &  16  & d \\
9824.13     &  [C I]      &  1F     & 9824.53  &    1.618  &	0.283  &   9  &  \\
9850.26     &  [C I]      &  1F     & 9850.66  &    5.069  &	0.882  &   9  &  \\
9903.46	    &  C II	  &  17.02  & 9903.55  &    0.214  &    0.037  &  13  & d \\
9903.88	    &  C II	  &  17.02  &          &           &           &      &  \\
10027.70    &  He I       &  6/7    &10027.91  &    0.878  &	0.148  &   9  &  \\
10049.37    &  H I        &  P7     &10049.64  &   30.179  &	5.080  &   9  &  \\
10320.49    &  [S II]     &  3F     &10320.77  &    4.780  &	0.771  &   9  &  \\
10336.41    &  [S II]     &  3F     &10336.66  &    4.039  &	0.650  &   9  &  \\
10370.50    &  [S II]     &  3F     &10370.79  &    2.109  &	0.338  &   9  &  \\
\noalign{\vskip3pt} \noalign{\hrule} \noalign{\vskip3pt}
\end{tabular}

\begin{tabular}{l@{\hspace{2.8mm}}l@{\hspace{2.8mm}}c@{\hspace{1.8mm}}
c@{\hspace{2.8mm}}c@{\hspace{2.8mm}}c@{\hspace{2.8mm}}r@{\hspace{1.8mm}}l@{\hspace{1.8mm}}}
\noalign{\hrule} \noalign{\vskip3pt}
$\lambda_{\rm 0}$& & & $\lambda_{\rm obs}$& & & err & Notes \\
(${\rm \AA}$)& Ion& Mult.& (${\rm \AA}$)& $F(\lambda)$& $I(\lambda)$$^{\rm a}$& (\%)$^{\rm b}$ & \\
\noalign{\vskip3pt} \noalign{\hrule} \noalign{\vskip3pt}
\multicolumn{8}{c}{\bf \large{M 20}}\\
\noalign{\vskip3pt} \noalign{\hrule} \noalign{\vskip3pt}
3187.84 &    He I      & 3     & 3187.72  &    2.364  &    3.375  &   5  &  \\
3354.42 &    He I      & 8     & 3354.54  &    0.177  &    0.239  &  21  &  \\
3447.59 &    He I      & 7     & 3447.54  &    0.280  &    0.370  &  15  &  \\
3478.97 &    He I      & 48    & 3479.09  &    0.087  &    0.115  &  35  &  \\
3487.73 &    He I      & 42    & 3487.76  &    0.064  &    0.083  &  :  &  \\
\end{tabular}
\end{minipage}
\end{table}

\setcounter{table}{1}
\begin{table}
\begin{minipage}{75mm}
\centering \caption{continued}
\begin{tabular}{l@{\hspace{2.8mm}}l@{\hspace{2.8mm}}c@{\hspace{1.8mm}}
c@{\hspace{2.8mm}}c@{\hspace{2.8mm}}c@{\hspace{2.8mm}}r@{\hspace{1.8mm}}l@{\hspace{1.8mm}}}
\noalign{\hrule} \noalign{\vskip3pt}
$\lambda_{\rm 0}$& & & $\lambda_{\rm obs}$& & & err & Notes \\
(${\rm \AA}$)& Ion& Mult.& (${\rm \AA}$)& $F(\lambda)$& $I(\lambda)$$^{\rm a}$& (\%)$^{\rm b}$ & \\
\noalign{\vskip3pt} \noalign{\hrule} \noalign{\vskip3pt}
3721.83 &    [S III]   & 2F    & 3721.68  &    0.914  &    1.130  &   6  &  \\
3721.94 &    H I       & H1    & 3721.92  &	 1.388  &    1.717  &	5  &  \\
3726.03 &    [O II]    & 1F    & 3725.97  &  118.617  &  146.604  &   3  &  \\
3728.82 &    [O II]    & 1F    & 3728.73  &  140.298  &  173.307  &   3  &  \\
3734.37 &    H I       & H13   & 3734.32  &    2.183  &    2.694  &   4  &  \\
3750.15 &    H I       & H12   & 3750.10  &    2.797  &    3.441  &   4  &  \\
\end{tabular}
\end{minipage}
\end{table}

\setcounter{table}{1}
\begin{table}
\begin{minipage}{75mm}
\centering \caption{continued}
\begin{tabular}{l@{\hspace{2.8mm}}l@{\hspace{2.8mm}}c@{\hspace{1.8mm}}
c@{\hspace{2.8mm}}c@{\hspace{2.8mm}}c@{\hspace{2.8mm}}r@{\hspace{1.8mm}}l@{\hspace{1.8mm}}}
\noalign{\hrule} \noalign{\vskip3pt}
$\lambda_{\rm 0}$& & & $\lambda_{\rm obs}$& & & err & Notes \\
(${\rm \AA}$)& Ion& Mult.& (${\rm \AA}$)& $F(\lambda)$& $I(\lambda)$$^{\rm a}$& (\%)$^{\rm b}$ & \\
\noalign{\vskip3pt} \noalign{\hrule} \noalign{\vskip3pt}
4658.10 & [Fe III] & 3F      & 4658.05  &    0.227  &    0.236  &   8  &  \\
4661.63 &   O II   & 1       & 4661.71  &    0.017  &    0.018  &  :  &  \\
4701.62 & [Fe III] & 3F      & 4701.50  &    0.050  &	 0.051  &  30  &  \\
4711.37 & [Ar IV]  & 1F      & 4711.50  &    0.030  &	 0.031  &  :  &  \\
4713.14 &    He I  & 12      & 4713.12  &    0.306  &    0.315  &   6  &  \\
4754.83 & [Fe III] & 3F      & 4754.69  &    0.028  &    0.029  &  :  &  \\
\end{tabular}
\end{minipage}
\end{table}

\setcounter{table}{1}
\begin{table}
\begin{minipage}{75mm}
\centering \caption{continued}
\begin{tabular}{l@{\hspace{2.8mm}}l@{\hspace{2.8mm}}c@{\hspace{1.8mm}}
c@{\hspace{2.8mm}}c@{\hspace{2.8mm}}c@{\hspace{2.8mm}}r@{\hspace{1.8mm}}l@{\hspace{1.8mm}}}
\noalign{\hrule} \noalign{\vskip3pt}
$\lambda_{\rm 0}$& & & $\lambda_{\rm obs}$& & & err & Notes \\
(${\rm \AA}$)& Ion& Mult.& (${\rm \AA}$)& $F(\lambda)$& $I(\lambda)$$^{\rm a}$& (\%)$^{\rm b}$ & \\
\noalign{\vskip3pt} \noalign{\hrule} \noalign{\vskip3pt}
7281.35 &  He I    & 45      & 7281.26  &    0.730  &	 0.522  &   5  &  \\
7298.05 &  He I    & 1/9     & 7298.00  &    0.040  &	 0.029  &  20  &  \\
7318.39 &  [O II]  & 2F      & 7318.88  &    0.763  &	0.543   &   5  &  \\
7319.99 &  [O II]  & 2F      & 7319.97  &    2.634  &	1.876   &   4  &  \\
7329.66 &  [O II]  & 2F      & 7329.61  &    1.882  &	1.339   &   4  &  \\
7330.73 &  [O II]  & 2F      & 7330.61  &    1.411  &	1.004   &   4  &  \\
\end{tabular}
\end{minipage}
\end{table}

\setcounter{table}{1}
\begin{table}
\begin{minipage}{75mm}
\centering \caption{continued}
\begin{tabular}{l@{\hspace{2.8mm}}l@{\hspace{2.8mm}}c@{\hspace{1.8mm}}
c@{\hspace{2.8mm}}c@{\hspace{2.8mm}}c@{\hspace{2.8mm}}r@{\hspace{1.8mm}}l@{\hspace{1.8mm}}}
\noalign{\hrule} \noalign{\vskip3pt}
$\lambda_{\rm 0}$& & & $\lambda_{\rm obs}$& & & err & Notes \\
(${\rm \AA}$)& Ion& Mult.& (${\rm \AA}$)& $F(\lambda)$& $I(\lambda)$$^{\rm a}$& (\%)$^{\rm b}$ & \\
\noalign{\vskip3pt} \noalign{\hrule} \noalign{\vskip3pt}
8518.04 & He I     & 2/8     & 8517.87  &    0.014  &    0.009  &   :  &  \\
8528.99 & He I     & 6/15    & 8528.99  &    0.025  &    0.016  &  31  &  \\
8665.02 & H I	   & P13     & 8664.93  &    2.131  &	 1.348  &   5  & c \\
8680.28 & N I	   &  1      & 8679.94  &    0.044  &	 0.028  &  19  &  \\
8683.40 & N I	   &  1      & 8685.83  &    0.027  &	 0.017  &  29  &  \\
8703.25 & N I	   &  1      & 8702.94  &    0.025  &	 0.016  &  30  &  \\
\end{tabular}

\begin{tabular}{l@{\hspace{2.8mm}}l@{\hspace{2.8mm}}c@{\hspace{1.8mm}}
c@{\hspace{2.8mm}}c@{\hspace{2.8mm}}c@{\hspace{2.8mm}}r@{\hspace{1.8mm}}l@{\hspace{1.8mm}}}
\noalign{\hrule} \noalign{\vskip3pt}
$\lambda_{\rm 0}$& & & $\lambda_{\rm obs}$& & & err & Notes \\
(${\rm \AA}$)& Ion& Mult.& (${\rm \AA}$)& $F(\lambda)$& $I(\lambda)$$^{\rm a}$& (\%)$^{\rm b}$ & \\
\noalign{\vskip3pt} \noalign{\hrule} \noalign{\vskip3pt}
\multicolumn{8}{c}{\bf \large{NGC 3603}}\\
\noalign{\vskip3pt} \noalign{\hrule} \noalign{\vskip3pt}
3686.83 & H I      & H19     & 3686.87  &    0.280  &    1.195  &  :  &  \\
3691.56 & H I      & H18     & 3691.73  &    0.309  &    1.311  &  37  &  \\
3697.15 & H I      & H17     & 3697.36  &    0.329  &    1.382  &  35  &  \\
3703.86 & H I      & H16     & 3704.04  &    0.360  &    1.485  &  32  &  \\
3705.04 & He I     & 25      & 3705.22  &    0.140  &    0.576  &  :  &  \\
3711.97 & H I      & H15     & 3712.24  &    0.346  &    1.411  &  33  &  \\
3721.83 &  [S III] & 2F      & 3722.00  &    0.774  &    3.119  &  17  &  \\
3721.94 &  H I     & H14     & 	        &	    &  	        &     &  \\
3726.03 &  [OII]   & 1F      & 3726.26  &   10.070  &	40.360  &   5  &  \\
3728.82 &  [OII]   & 1F      & 3728.95  &    6.256  &	24.990  &   5  &  \\
\noalign{\vskip3pt} \noalign{\hrule} \noalign{\vskip3pt}
\end{tabular}
\end{minipage}
\end{table}

\setcounter{table}{1}
\begin{table}
\begin{minipage}{75mm}
\centering \caption{continued}
\begin{tabular}{l@{\hspace{2.8mm}}l@{\hspace{2.8mm}}c@{\hspace{1.8mm}}
c@{\hspace{2.8mm}}c@{\hspace{2.8mm}}c@{\hspace{2.8mm}}r@{\hspace{1.8mm}}l@{\hspace{1.8mm}}}
\noalign{\hrule} \noalign{\vskip3pt}
$\lambda_{\rm 0}$& & & $\lambda_{\rm obs}$& & & err & Notes \\
(${\rm \AA}$)& Ion& Mult.& (${\rm \AA}$)& $F(\lambda)$& $I(\lambda)$$^{\rm a}$& (\%)$^{\rm b}$ & \\
\noalign{\vskip3pt} \noalign{\hrule} \noalign{\vskip3pt}
3734.37 &   H I    & H13     & 3734.56  &    0.715  &	 2.835  &  18  &  \\
3750.15 &   H I    & H12     & 3750.43  &    1.017  &	 3.951  &  14  &  \\
3770.63 &   H I    & H11     & 3770.84  &    1.158  &	 4.385  &  12  &  \\
3797.90 &   H I    & H10     & 3798.09  &    1.517  &	 5.553  &  10  &  \\
3819.61 &   He I   & 20      & 3819.83  &    0.379  &	 1.352  &  31  &  \\
3835.39 & H I      & H9      & 3835.58  &    2.344  &	 8.205  &   7  &  \\
\end{tabular}
\end{minipage}
\end{table}

\setcounter{table}{1}
\begin{table}
\begin{minipage}{75mm}
\centering \caption{continued}
\begin{tabular}{l@{\hspace{2.8mm}}l@{\hspace{2.8mm}}c@{\hspace{1.8mm}}
c@{\hspace{2.8mm}}c@{\hspace{2.8mm}}c@{\hspace{2.8mm}}r@{\hspace{1.8mm}}l@{\hspace{1.8mm}}}
\noalign{\hrule} \noalign{\vskip3pt}
$\lambda_{\rm 0}$& & & $\lambda_{\rm obs}$& & & err & Notes \\
(${\rm \AA}$)& Ion& Mult.& (${\rm \AA}$)& $F(\lambda)$& $I(\lambda)$$^{\rm a}$& (\%)$^{\rm b}$ & \\
\noalign{\vskip3pt} \noalign{\hrule} \noalign{\vskip3pt}
4931.32 & [O III]  & 1F      & 4931.45  &    0.045  &	 0.040  &  :  &  \\
4958.91 & [O III]  & 1F      & 4959.20  &  204.749  &  180.464  &   3  &  \\
4985.90 & [Fe III] & 2F      & 4985.75  &    0.164  &	 0.139  &  37  &  \\
4987.20 & [Fe III] & 2F      & 4987.85  &    0.280  &	 0.238  &  25  &  \\
5006.84 & [O III]  & 1F      & 5007.15  &  641.906  &  533.117  &   3  &  \\
5015.68 & He I     & 4       & 5015.99  &    2.663  &    2.188  &   5  &  \\
\end{tabular}
\end{minipage}
\end{table}

\setcounter{table}{1}
\begin{table}
\begin{minipage}{75mm}
\centering \caption{continued}
\begin{tabular}{l@{\hspace{2.8mm}}l@{\hspace{2.8mm}}c@{\hspace{1.8mm}}
c@{\hspace{2.8mm}}c@{\hspace{2.8mm}}c@{\hspace{2.8mm}}r@{\hspace{1.8mm}}l@{\hspace{1.8mm}}}
\noalign{\hrule} \noalign{\vskip3pt}
$\lambda_{\rm 0}$& & & $\lambda_{\rm obs}$& & & err & Notes \\
(${\rm \AA}$)& Ion& Mult.& (${\rm \AA}$)& $F(\lambda)$& $I(\lambda)$$^{\rm a}$& (\%)$^{\rm b}$ & \\
\noalign{\vskip3pt} \noalign{\hrule} \noalign{\vskip3pt}
8045.63 & [Cl IV]  & 1F      & 8045.92  &    1.042  &    0.080  &   7  &  \\
	& [Cl IV] & red c. & 8046.82  &    0.394  &	 0.030	&   9  & f \\
8057.   & He I     & 4/18    & 8058.05  &    0.142  &    0.011  &  13  &  \\
8116.   & He I     & 4/16    & 8116.98  &    0.169  &    0.013  &  12  &  \\
8155.66 & He I     & 4/15    & 8156.10  &    0.173  &    0.013  &  12  &  \\
8188.01 & N I      & 2       & 8188.55  &    0.196  &    0.014  &  11  &  \\
\end{tabular}
\end{minipage}
\end{table}

\setcounter{table}{1}
\begin{table}
\begin{minipage}{75mm}
\centering \caption{continued}
\begin{tabular}{l@{\hspace{2.8mm}}l@{\hspace{2.8mm}}c@{\hspace{1.8mm}}
c@{\hspace{2.8mm}}c@{\hspace{2.8mm}}c@{\hspace{2.8mm}}r@{\hspace{1.8mm}}l@{\hspace{1.8mm}}}
\noalign{\hrule} \noalign{\vskip3pt}
$\lambda_{\rm 0}$& & & $\lambda_{\rm obs}$& & & err & Notes \\
(${\rm \AA}$)& Ion& Mult.& (${\rm \AA}$)& $F(\lambda)$& $I(\lambda)$$^{\rm a}$& (\%)$^{\rm b}$ & \\
\noalign{\vskip3pt} \noalign{\hrule} \noalign{\vskip3pt}
8728.90 & [Fe III] & 8F      & 8729.93  &    0.166  &    0.008  &  13  &  \\
8728.90 & N I      & 28      &          &           &           &     &  \\
8733.43 & He I     & 6/12    & 8734.04  &    0.996  &	 0.047  &   8  &  \\
8736.04 & He I     & 7/12    & 8736.72  &    0.349  &	 0.016  &  10  &  \\
8750.47 & H I      & P12     & 8751.07  &   26.668  &	 1.234  &   8  &  \\
8776.77 & He I     & 4/9     & 8777.33  &    0.821  &    0.037  &   9  &  \\
\end{tabular}
\begin{description}
\item[$^{\rm a}$] c(H$\beta$), $I$(H$\beta$) values per nebula are: M16 (1.21, $7.331 \times 10^{-12}$ 
ergs cm$^{-2}$ s$^{-1}$); M20 (0.36, $1.081 \times 10^{-12}$ ergs cm$^{-2}$ s$^{-1}$) and NGC~3603 (2.36, 
$6.506 \times 10^{-11}$ ergs cm$^{-2}$ s$^{-1}$).
\item[$^{\rm b}$] Colons indicate uncertainties larger than 40 \%.
\item[$^{\rm c}$] Affected by telluric emission lines.
\item[$^{\rm d}$] Affected by atmospheric absorption bands.
\item[$^{\rm e}$] Affected by internal reflections or charge transfer in the CCD.
\item[$^{\rm f}$] Blend with an unknown line.
\item[$^{\rm g}$] Dubious identification.
\end{description}\end{minipage}
\end{table}

Table~\ref{lineid} presents the emission line intensities measured in the three {\hii} regions. The
first and fourth columns include the adopted laboratory wavelength, $\lambda_0$, and 
the observed wavelength in the heliocentric restframe, $\lambda_{\rm obs}$. 
The second and third columns include the ion and the multiplet number, or 
series for each line.  The fifth and sixth columns list the observed 
flux relative to H$\beta$, $F(\lambda$), and the reddening corrected flux 
relative to H$\beta$, $I(\lambda$). The seventh column includes the 
fractional error (1$\sigma$) in the line intensities. Errors were derived following
\citep{garciarojasetal04}, adding quadratically the error due to flux calibration that has been estimated to
be about 3\%, which corresponds to the standard deviation obtained from the calibration 
curves of the three standard stars.

A total of 256, 261 and 235 emission lines were measured in M16, M20 and NGC~3603, respectively. 
Most of the lines are permitted lines. We have measured 54 forbidden lines (CELs) in M16, 58 in M20 and 60 in NGC~3603. 
We have detected several semiforbidden lines: 8 in M16, 12 in M20, and only 2 in NGC~3603 (see Table~\ref{lineid}). 
Four lines detected in NGC~3603 were identified as red velocity components of highly ionised species (see 
\S~\ref{veloc} for a detailed discussion on these lines).
In several cases some identified lines were severely blended with telluric lines, making impossible 
their measurement. Other lines were strongly affected by atmospheric features in absorption 
or by internal reflections by charge transfer in the CCD, rendering their intensities unreliable. 
Also, several lines were labelled as dubious identifications and two emission lines 
could not be identified in any of the available references. All those lines are indicated in 
Table~\ref{lineid}.

The identification and adopted laboratory wavelengths of the lines were obtained following 
several previous identifications in the literature \citep[see][and references therein]{garciarojasetal04, 
estebanetal04}. Several identifications labelled as dubious in \citet{garciarojasetal04, 
garciarojasetal05} have been updated in this work following a criteria based in the comparison of the radial 
velocity of the line with lines of similar ionisation potential.

We have assumed the standard dust extinction law for the Milky Way (R$_{\rm v}$=3.1) parametrized by 
\citet{seaton79} for M20 and NGC~3603. 
A reddening coefficient of c(H$\beta$)=0.36$\pm$0.04 was determined 
for M20, by fitting the observed 
$I$(H Balmer lines)/$I$(H$\beta$) ratios (from H16 to H$\beta$) and $I$(H Paschen 
lines)/$I$(H$\beta$) (from P22 to P7), to the theoretical 
ones computed by Storey \& Hummer (1995) for {\te} = 10000 K and {\elecd} = 1000 cm$^{-3}$ (see 
below). {\hi} lines affected by blends or atmospheric absorption were not considered. 
Our derived c(H$\beta$) is slightly lower than previous determinations in M20, 
but it has been derived with a much larger number of {\hi} lines:
\citet{hawley78} derived c(H$\beta$)=0.42 and 0.48 for two slit positions with 
offsets of 33$\arcsec$ south, 10$\arcsec$ west, and  33$\arcsec$ south, 25$\arcsec$ east from 
our slit position; \citet{lyndsoneil85} computed a value of c(H$\beta$)=0.45 from long slit 
spectroscopic data for a larger extension of the nebula.

Following the same method as for M20, we have derived c(H$\beta$)=2.36$\pm$0.06 for NGC~3603. 
\citet{tapiaetal01} derived a c(H$\beta$)=2.51 for a slit position 116$\arcsec$ east and 12$\arcsec$ north 
from our slit position. \citet{melnicketal89} derived an average c(H$\beta$)=1.93 for four slit positions.
\citet{girardietal97} obtained c(H$\beta$)=2.36 and 2.59 for two slit positions in 
NGC~3603 located near ours and using the extinction law of \citet{savagemathis79}. Even though the 
different extinction laws used are different, these values of c(H$\beta$)
are in very good agreement with our value. Using the extinction 
law by \citet{savagemathis79} we have obtained c(H$\beta$)=2.29$\pm$0.06 which is consistent with 
our adopted value. 

On the other hand, \citet{chiniwargau90} confirmed an abnormal extinction of dense interstellar dust 
within M16 from photometric observations of the associated young stellar cluster NGC~6611. They 
found that deviations from the normal extinction law occur at wavelenghts shorter 
than 5500 ${\rm \AA}$, because of the higher size of the graphite grains.
Following those authors we have assumed the extinction law parametrized by \citet{cardellietal89} 
with the ratio of total to selective extinction, R$_{\rm v}$=3.1 for wavelengths higher than 
5500 ${\rm \AA}$, and R$_v$=4.8 for shorter wavelengths. From this extinction law and following 
the same procedure than for M20 and NGC~3603, we have derived c(H$\beta$)=1.21$\pm$0.06 for M16, assuming 
{\te} = 8000 K and {\elecd} = 1000 cm$^{-3}$ \citep{rodriguez98}.

\section{Physical Conditions}\label{phiscond}

\subsection{Temperatures and Densities}

The large number of emission lines identified and measured in the spectra allows us to derive physical 
conditions using different emission
line ratios. The temperatures and densities are presented in Table~\ref{plasma}. Most of the 
determinations were carried out with the 
{\sc IRAF} task {\sc TEMDEN} of the package {\sc NEBULAR} \citep{shawdufour95}. 

The methodology followed for the derivation of {\elecd} and {\te}, and the atomic data 
compilation used have been described in previous papers  
\citep[i.e.][]{garciarojasetal04,garciarojasetal05}. In the case of electron densities, ratios of CELs 
of several ions have been used.
We have derived the {\ffeiii} density from the intensity of the brightest 
lines (which have errors equal or smaller than 30 \% and that seem not to be affected by line blending, see Table~\ref{lineid})  
together with the computations of \citet{rodriguez02}. We have used 4, 6 and 4 {\ffeiii} lines 
for M16, M20 and NGC~3603 respectively; the methodology consisted of adopting the density that 
minimized the dispersion of individual Fe$^{++}$/H$^+$ abundance ratios. All the computed values of {\elecd} are consistent 
within the errors (see Table~\ref{plasma}).

\setcounter{table}{2}
\begin{table*}
\begin{minipage}{150mm}
\centering \caption{Plasma Diagnostic.}
\label{plasma}
\begin{tabular}{lllll}
\noalign{\hrule} \noalign{\vskip3pt}
Parameter & Line & \mc{3}{c}{Value} \\
\noalign{\vskip3pt} \noalign{\hrule} \noalign{\vskip3pt}
& & \mc{1}{c}{M16}& \mc{1}{c}{M20} & \mc{1}{c}{NGC 3603} \\
\noalign{\vskip3pt} \noalign{\hrule} \noalign{\vskip3pt}
{\elecd} (cm$^{-3}$)& {\fni} ($\lambda$5198)/($\lambda$5200)& 1100$^{+750}_{-400}$& 560$^{+340}_{-220}$ & 4000: \\
& {\foii} ($\lambda$3726)/($\lambda$3729)& 1050 $\pm$ 250& 240 $\pm$ 70 & 2300 $\pm$ 750 \\
& {\foii} ($\lambda$3726+$\lambda$3729)/($\lambda$7320+$\lambda$7330)$^{\rm a}$&  &  & 5300$\pm$850 \\
& {\fsii} ($\lambda$6716)/($\lambda$6731)& 1390$\pm$550& 320$\pm$130 & 4150$^{+3350}_{-1650}$  \\ 
& {\ffeiii} &  540$^{+>1000}_{-500}$& 560$\pm$390 & 1330: \\
& {\fcliii} ($\lambda$5518)/($\lambda$5538)& 1370$\pm$1000& 350$^{+780}_{-350}$ &  5600$^{+3900}_{-2400}$ \\ 
& {\fariv} ($\lambda$4711)/($\lambda$4740) & \mc{1}{c}{---}  & \mc{1}{c}{---} & $\leq 4000$  \\
& {\elecd} (adopted) & 1120$\pm$220& 270$\pm$60 & 5150$\pm$750  \\
& & \\
T$_{\rm e}$ (K)& {\fnii} ($\lambda$6548+$\lambda$6583)/($\lambda$5755)$^{\rm a}$& 8450 $\pm$ 270 & 8500$\pm$240 & 11050$\pm$800  \\
& {\fsii} ($\lambda$6716+$\lambda$6731)/($\lambda$4069+$\lambda$4076)&  7520 $\pm$430& 6950 $\pm$350 &  11050 $^{+3550}_{-2050}$  \\
& {\foii} ($\lambda$3726+$\lambda$3729)/($\lambda$7320+$\lambda$7330)$^{\rm a}$& 8260$\pm$400& 8275$\pm$400  &  12350$\pm$1250  \\
& T$_{\rm e}$ (low) & 8350$\pm$200& 8400$\pm$200 &  11400$\pm$700 \\
& {\foiii} ($\lambda$4959+$\lambda$5007)/($\lambda$4363) & 7650 $\pm$ 250& 7800 $\pm$ 300 & 9060 $\pm$200  \\
& {\fariii} ($\lambda$7136+$\lambda$7751)/($\lambda$5192)& \mc{1}{c}{---} & 8730 $\pm$920$^{\rm b}$  & \mc{1}{c}{---}  \\
& {\fsiii} ($\lambda$9069+$\lambda$9532)/($\lambda$6312) & 8430 $\pm$450& 8300 $\pm$400  & 8800 $\pm$500 \\
& T$_{\rm e}$ (high) & 7850$\pm$220& 7980$\pm$250 &  9030$\pm$200 \\
& {\hei} & 7300$\pm$350& 7650$\pm$300 &  8480$\pm$200 \\
& Balmer decrement & 5450$\pm$820& 6000$\pm$1300 &  \mc{1}{c}{---} \\ 
& Paschen decrement & 7200$\pm$1300& 5700$\pm$1300 & 6900$\pm$1100  \\ 
\noalign{\vskip3pt} \noalign{\hrule} \noalign{\vskip3pt}
\end{tabular}
\begin{description}
\item[$^{\rm a}$] The recombination contribution to the auroral lines has been taken into account (see text)
\item[$^{\rm b}$] The {\fariii} $\lambda$7751 line is severaly blended with a telluric line.
\end{description}\end{minipage}
\end{table*}

A weighted mean of {\elecd}({\oii}), {\elecd}({\feiii}), {\elecd}({\cliii}) and 
{\elecd}({\sii}) has been used to derive {\te}({\nii}), {\te}({\oii}), {\te}({\sii}), 
{\te}({\oiii}), {\te}({\ariii}) and {\te}({\siii}), and we iterated until convergence. For M20, 
which has a low ionisation degree, this is the first time that it has been possible to derive 
temperatures associated with high-ionised species. The values adopted for
{\elecd} are shown in Table~\ref{plasma}. 
We have excluded {\elecd}({\nitroi}) from the average because this ion is representative of the 
very outer part of the nebula, and does not coexist with the other ions.

Electron temperatures have been derived from the ratio of CELs of several ions and making use of 
{\sc NEBULAR} routines with upgraded atomic data for {\fsiii} \citep[see][]{garciarojasetal05}. 

We have corrected {\te}({\oii}) for the contribution to $\lambda\lambda$7320+7330 due to 
recombination following the formula derived by \citet{liuetal01} (their equation 2). 
From our {\oii} recombination lines we have estimated contributions of about 3\%, 2\% and 7\% for M16, M20 and 
NGC~3603, respectively.

Also, the contribution to the intensity of the {\fnii} $\lambda$5755 
line due to recombination can be estimated from an expression given by 
\citet{liuetal01} (their equation 1). From our data, 
we have obtained a contribution of recombination of
about 2\%, for M16 and NGC~3603, that does not affect significantly the temperature 
determination. For M20, the derived contribution was less than 0.1 \%, which is absolutely 
negligible\footnote{Although the formation mechanism of {\nii} permitted lines is mostly resonance 
fluorescence by the recombination line {\hei} $\lambda$508.64 ${\rm \AA}$ \citep{grandi76}, 
we have estimated N$^{++}$/H$^{+}$ abundances from {\nii} lines of multiplet 3, which is 
less affected by such effects. Anyway, the correction is in all the cases very small, 
and effects due to resonance fluorescence do not modify the derived temperature 
by more than 100 K.}.

\begin{figure*}
\begin{center}
\epsfig{file=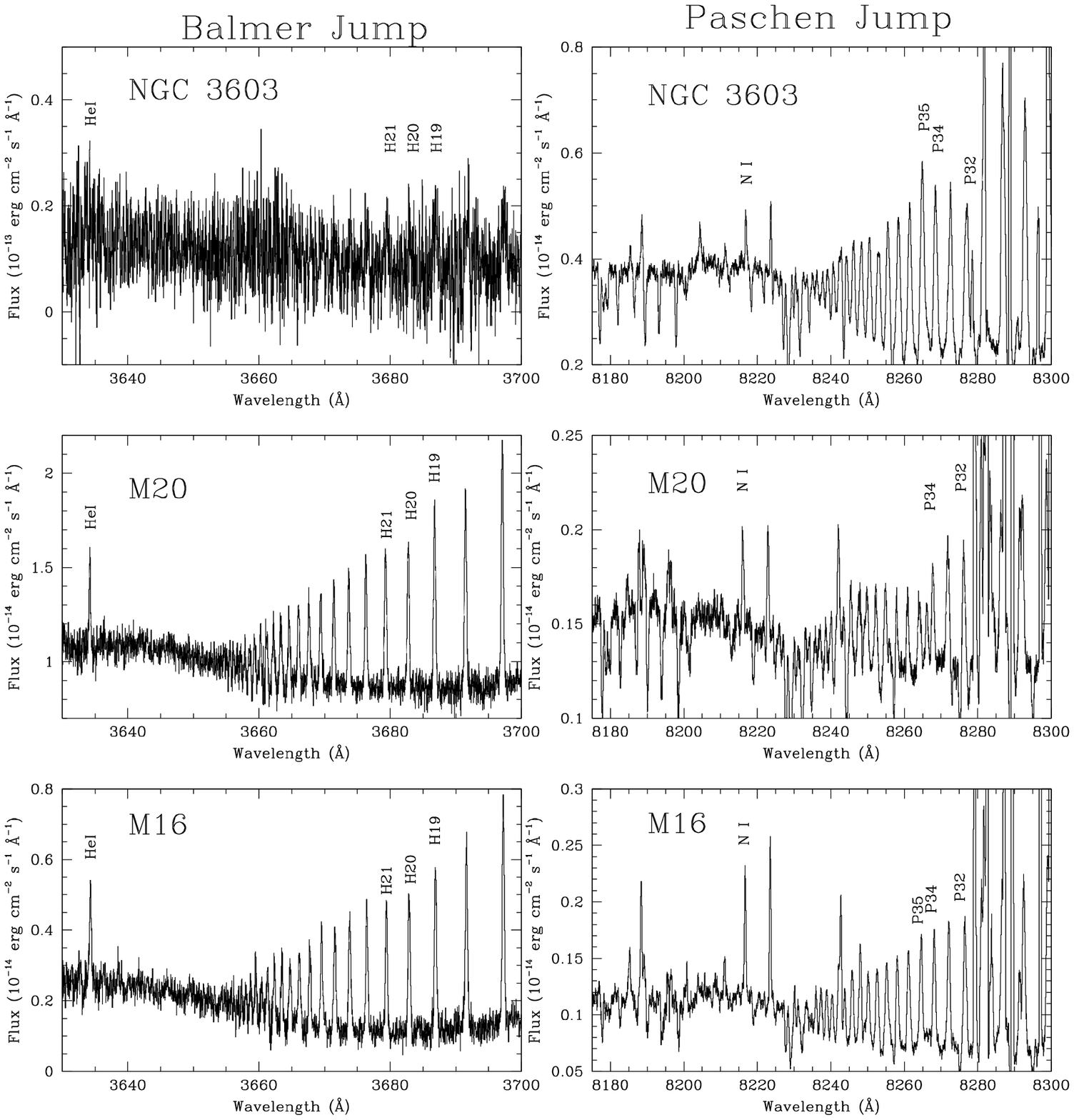, width=15. cm, clip=}22
\caption{Section of the echelle spectra of the three nebulae including the Balmer (left) and the Paschen (right) 
limits (observed fluxes).
\label{saltos}}
\end{center}
\end{figure*}

Figure~\ref{saltos} shows the spectral regions near the Balmer and the Paschen limits. 
The discontinuities can be clearly appreciated, except in the case of the Balmer limit in NGC~3603. 
Discontinuities are defined as
$I_c(Bac)=I_c(\lambda3646^-)-I_c(\lambda3646^+)$ and 
$I_c(Pac)=I_c(\lambda8203^-)-I_c(\lambda8203^+)$ respectively. 
The high spectral resolution of the spectra permits to measure the continuum emission in zones very near de 
discontinuity, minimizing the possible contamination of other continuum 
contributions. In fact, on the blue part of Paschen discontinuity, we have selected a zone that 
is free of contamination by telluric absorption. That zone is between 8203 and 8209 
{\rm \AA}. The uncertainty in the derived continua is the standard mean deviation of the averaged continua. On 
the red part, the measurement of the continuum is much easier, and could be computed as an average of the continua between 
the closest Paschen lines. 
We have obtained power-law fits to the relation between 
$I_c(Bac)/I(Hn)$ or $I_c(Pac)/I(Pn)$ and {\te} for different $n$ corresponding to 
different observed lines of both series. The emissivities as a function of electron 
temperature for the nebular continuum and the {\hi} Balmer and Paschen lines 
have been taken from \citet{brownmathews70} and \citet{storeyhummer95},  respectively. The $T_e(Bac)$ 
adopted is the average of the values using the lines from $H\alpha$ to 
H~10 (the brightest ones). In the case of $T_e(Pac)$, the adopted value is the average 
of the individual temperatures obtained using the lines from P~7 to P~13 (the brightest 
lines of the series), excluding those lines whose intensity seems to be 
affected by telluric lines or sky absorption. To our knowledge this is the first time 
that Balmer and Paschen temperatures have been derived for these three nebulae. 

We have compared our averaged {\te}({\hi}) with those derived from radio {\hi} recombination lines. 
\citet{caswellhaynes87} obtained {\te}({\hi}) = 6900 K for NGC~3603, the same value we derive for the Paschen 
temperature of this object; \citet{reifensteinetal70} derived {\te}({\hi}) = 6100 $\pm$ 1500 K for M16 from the H109$\alpha$
radio recombination line, which is also in good agreement with our average temperature; however, for M20, 
\citet{reifensteinetal70} derived {\te}({\hi}) = 7300 $\pm$ 2500, which is higher than the {\te}({\hi}) we 
obtain, although it is compatible within the errors. Effects of scattered continuum 
light in the continua of Balmer and Paschen limits may be a reason of this discrepancy (see \S~\ref{contm20}). 
However, aperture effects cannot be ruled out. Radio determinations usually refer to 
average values over large extensions of the nebula, instead our values refer to very small and particular 
zones of the nebulae. Another interesting comparison is with 
{\te}({\hi}) derived through radio continuum measurements. \citet{shavergoss70} estimate 
{\te}({\hi}) = 6800 $\pm$ 700 K for NGC~3603, from 408 Mhz continuum measurements, which is in excellent 
agreement with the estimate through radio {\hi} recombination lines and our own spectroscopic value.

\citet{apeimbertetal02} developed a method to derive the helium temperature, {\te}({\hei}), in the 
presence of temperature fluctuations. Assuming a 2-zone ionization scheme and the formulation 
of \citet{apeimbertetal02} we have derived {\te}({\hei})=7300 $\pm$ 350 K, 7650 $\pm$ 300 and 8480 $\pm$ 200 
for M16, M20 and NGC~3603, respectively. These results are higher than those derived from {\hi}.

We have assumed a 2-zone ionisation scheme for the calculation of ionic abundances 
(see \S~\ref{abundances}). We have adopted the average of electron temperatures 
obtained from {\fnii} and {\foii} lines as representative for the low ionisation zone designated {\te}(low). 
The average of electron temperatures obtained from {\foiii} and {\fsiii} lines has been 
assumed as representative  of the high ionisation zone designated {\te}(high) (see Table~\ref{plasma}).

\subsection{Temperature variations}\label{tsq}

\citet{torrespeimbertetal80} proposed the
presence of spatial temperature fluctuations (parametrized by {\ts}) as the cause of the
discrepancy between abundance calculations based on CELs and RLs. This is due to the different dependence 
on the electron temperature of the CELs and RLs emissivities. 
Assuming the validity of the temperature fluctuations paradigm, the comparison of the 
abundances determined from both kinds of lines for a given ion should provide an estimation of {\ts}. 
Also, \citet{peimbert71} proposed that there is a dichotomy between {\te} derived from 
the {\foiii} lines and from the hydrogen recombination continuum discontinuities, which is
correlated with the discrepancy between CEL and RL abundances \citep[e.g.][]{peimbertcostero69, 
torrespeimbertetal80, liuetal00, tsamisetal04}, 
so the comparison between electron temperatures obtained from both methods is an additional
indicator of {\ts}. 
A complete formulation of temperature fluctuations has been developed by \citet{peimbert67},
\citet{peimbertcostero69} and \citet{peimbert71} \citep[see also][]{apeimbertetal02, ruizetal03}. 
\citet{esteban02} discussed some of the different mechanisms 
proposed to explain the presence of temperature fluctuations in nebulae, it is beyond the scope of the 
present paper to treat this topic.

As we have assumed a two-zone ionisation scheme, we have followed 
the re-formulation proposed by \citet{peimbertetal00} and \citet{apeimbertetal02} to derive the value of {\ts} 
comparing the average of {\te}($Bac$) 
and {\te}($Pac$) with the combination of {\te}({\foii}) and {\te}({\foiii}), $T_e$({\oii+\iii}), using equation (A1) of 
\citet{apeimbertetal02}. In Table~\ref{t2} we include the different {\ts} values that produce the agreement 
between the abundance determinations obtained from CELs and RLs of O$^+$ (for those objects where 
{\oi} RLs have been measured) and O$^{++}$, 
as well as the values of {\ts} obtained from the combination of $T_e$({\oii+\iii}) and the average value of 
{\te}($Bac$) and {\te}($Pac$). As it can be seen, the different {\ts} values obtained are rather consistent. 
In Table~\ref{t2}, we also include the {\ts} value obtained from the application of a 
maximum likelihood method to search for the physical conditions, including He$^+$/H$^+$ ratios and optical depths, 
that would be a simultaneous fit to all the measured lines of {\hei} (see \S~\ref{heabund}). 
Finally, in Table~\ref{t2} we show the final adopted values, which are error-weighted averages.

\setcounter{table}{3}
\begin{table}
\begin{minipage}{75mm}
\centering \caption{{\ts} parameter}
\label{t2}
\begin{tabular}{cccc}
\noalign{\hrule} \noalign{\vskip3pt}
Method & \mc{3}{c}{\ts} \\
\noalign{\vskip3pt} \noalign{\hrule} \noalign{\vskip3pt}
& \mc{1}{c}{M16}& \mc{1}{c}{M20} & \mc{1}{c}{NGC 3603} \\
\noalign{\vskip3pt} \noalign{\hrule} \noalign{\vskip3pt}
O$^{\rm ++}$ (R/C)&  0.046$\pm$0.007&  0.038$\pm$0.016& 0.042$\pm$0.009 \\
O$^{\rm +}$ (R/C)& --- & 0.032$\pm$0.020& --- \\
He$^{\rm +}$ & 0.017$\pm$0.013& 0.017$\pm$0.010& 0.032$\pm$0.014 \\
Bac/Pac--FL & 0.045$\pm$0.014& 0.049$\pm$0.019& 0.056$\pm$0.023 \\ 
Adopted & 0.039$\pm$0.006& 0.029$\pm$0.007 &  0.040$\pm$0.008 \\ 
\noalign{\vskip3pt} \noalign{\hrule} \noalign{\vskip3pt}
\end{tabular}
\end{minipage}
\end{table}

\section{Ionic Abundances}\label{abundances}

\subsection{He$^{+}$ abundance}\label{heabund}

\setcounter{table}{4}
\begin{table}
\begin{minipage}{75mm}
\centering \caption{He$^+$ abundance.}
\label{he}
\begin{tabular}{lcccc}
\noalign{\hrule} \noalign{\vskip3pt}
Line & \multicolumn{3}{c}{He$^+$/H$^+$ $^{\rm a}$} \\
\noalign{\vskip3pt} \noalign{\hrule} \noalign{\vskip3pt}
& \mc{1}{c}{M16}& \mc{1}{c}{M20} & \mc{1}{c}{NGC 3603} \\
\noalign{\vskip3pt} \noalign{\hrule} \noalign{\vskip3pt}
3819.61& 827 $\pm$ 50 	& 731 $\pm$ 44 & 1014 $\pm$ 314 \\
3888.65& ---	 	& 695 $\pm$ 21 & --- 		\\
3964.73& 797 $\pm$ 32 	& 749 $\pm$ 30 & 820 $\pm$ 90 \\ 
4026.21& 834 $\pm$ 25 	& 756 $\pm$ 22 & 1043 $\pm$ 63 \\ 
4387.93& 777 $\pm$ 31 	& 678 $\pm$ 41 & 1067 $\pm$ 107 \\ 
4471.09& 758 $\pm$ 23 	& 690 $\pm$ 21 & 998 $\pm$ 30 \\
4713.14& 794 $\pm$ 32 	& 674 $\pm$ 40 & 1001 $\pm$ 70 \\
4921.93& 757 $\pm$ 23 	& 723 $\pm$ 22 & 926 $\pm$ 37 \\ 
5875.64& 761 $\pm$ 23 	& 672 $\pm$ 19 & 921 $\pm$ 37 \\ 
6678.15& 751 $\pm$ 30 	& 702 $\pm$ 28 & 911 $\pm$ 46 \\
7065.28& 781 $\pm$ 39 	& 703 $\pm$ 28 & 956 $\pm$ 57 \\ 
7281.35& 831 $\pm$ 41 	& 787 $\pm$ 39 & 957 $\pm$ 57 \\ \hline
Adopted$^{\rm b}$& 781 $\pm$ 12 & 711 $\pm$ 10 & 961 $\pm$ 17 \\ 
\noalign{\vskip3pt} \noalign{\hrule} \noalign{\vskip3pt}
\end{tabular}
\begin{description}
\item[$^{\rm a}$] In units of 10$^{-4}$, for $\tau_{3889}$= 2.99 $\pm$ 0.85, 2.09 $\pm$ 
0.49, and 12.12 $\pm$ 1.00, and {\ts}=0.039$\pm$0.006, 0.029$\pm$0.007 and 0.040$\pm$0.008
Uncertainties correspond to line intensity errors.
\item[$^{\rm b}$] It includes all the relevant uncertainties in emission line intensities, 
{\elecd}, $\tau_{3889}$ and {\ts}.
\end{description}
\end{minipage}
\end{table}

We have measured 47, 53 and 64 {\hei} emission lines in the spectra of M16, 
M20 and NGC~3603, respectively. 
These lines arise mainly from recombination but they can be affected by collisional excitation and 
self-absorption effects. We have
determined the He$^+$/H$^+$ ratio from a maximum likelihood method \citep[e.g.][]{peimbertetal00},
using the {\elecd} given in Table~\ref{plasma} and $T$({\oii}+{\sc iii})=8130 K for M16, 
$T$({\oii}+{\sc iii})=8200 K for M20 and $T$({\oii}+{\sc iii})=9600 K for NGC~3603 
(see \S~\ref{tsq}).
We have used the effective recombination coefficients of \citet{storeyhummer95} 
for {\hi} and those of \citet{smits96} and \citet{benjaminetal99} for {\hei}. 
The collisional contribution
was estimated from \citet{saweyberrington93} and \citet{kingdonferland95}, and the optical depth in 
the triplet lines were
derived from the computations by \citet{benjaminetal02}. 

In Table~\ref{he} we have included the He$^+$/H$^+$ ratios we have obtained for the individual 
{\hei} lines not affected by line blending and with the highest signal-to-noise ratio. We have
excluded {\hei} $\lambda$5015 because it could suffer self-absorption effects from the 2$^1$S metastable level, 
as was already pointed out by \citet{estebanetal04}. We have also excluded  
$\lambda$3889 for M16 and NGC3603 because it is severely blended with the Balmer H8 line. 
We have performed a $\chi^2$ 
optimisation of the values given in the table, and we have obtained a $\chi^2$ parameter of 8.3, 15.1 and 
9.63 for M16, M20 and NGC~3603, respectively, these values indicate a reasonable goodness of the fits for a system 
with nine degrees of freedom.

\subsection{Ionic Abundances from CELs}\label{cels}

Ionic abundances of N$^+$, O$^+$, O$^{++}$, Ne$^{++}$, S$^+$, S$^{++}$, Cl$^+$, Cl$^{++}$, Cl$^{3+}$
Ar$^{++}$ and Ar$^{3+}$ have been determined from CELs, using the {\sc IRAF} package {\sc NEBULAR} 
except for Cl$^+$ \citep[see][]{garciarojasetal04}. 
Additionally, we have determined the ionic abundances of Fe$^{++}$, which we will discuss further on. Ionic abundances are listed in 
Table~\ref{celabun} and correspond to the mean value of the 
abundances derived from all the individual lines of each ion observed (weighted by their relative strengths).  

To derive the abundances for $t^2$$>$0.00 (see \S~\ref{t2}) we used the abundances for $t^2$=0.00 and the 
formulation of by \citet{peimbert67} and \citet{peimbertcostero69}. 
For other $t^2$ values, it is possible to interpolate or extrapolate 
the values presented in Table~\ref{celabun}.

\setcounter{table}{5}
\begin{table*}
\begin{minipage}{150mm}
\centering \caption{Ionic abundances from collisionally excited lines$^{\rm a}$.}
\label{celabun}
\begin{tabular}{lcccccc}
\noalign{\hrule} \noalign{\vskip3pt}
& \mc{2}{c}{M16}& \mc{2}{c}{M20} & \mc{2}{c}{NGC 3603} \\
\noalign{\vskip3pt} \noalign{\hrule} \noalign{\vskip3pt}
Ion & {\ts}=0.000 & {\ts}=0.039$\pm$0.006 & {\ts}=0.000 & {\ts}=0.029$\pm$0.007 &  {\ts}=0.000 & {\ts}=0.040$\pm$0.008 \\
\noalign{\vskip3pt} \noalign{\hrule} \noalign{\vskip3pt}
N$^{0}$	 &6.15$\pm$0.06	& 6.33$\pm$0.07 & 5.90$\pm$0.07 & 6.03$\pm$0.08& 5.65$\pm$0.11 	& 5.75$\pm$0.11 \\
N$^{+}$	 &7.71$\pm$0.05 & 7.88$\pm$0.06 & 7.55$\pm$0.04 & 7.67$\pm$0.05& 6.45$\pm$0.07 	& 6.55$\pm$0.07 \\
O$^{0}$	 &7.23$\pm$0.05 & 7.40$\pm$0.06 & 6.60$\pm$0.05 & 6.72$\pm$0.06& 6.32$\pm$0.09 	& 6.42$\pm$0.09 \\
O$^{+}$	 &8.47$\pm$0.08 & 8.66$\pm$0.09 & 8.46$\pm$0.07 & 8.59$\pm$0.08& 7.44$\pm$0.11 	& 7.54$\pm$0.11 \\
O$^{++}$ & 7.85$\pm$0.07& 8.18$\pm$0.10 & 7.67$\pm$0.08 & 7.90$\pm$0.10& 8.42$\pm$0.05 	& 8.68$\pm$0.08 \\
Ne$^{++}$& 7.01$\pm$0.07& 7.38$\pm$0.10 & 6.55$\pm$0.09 & 6.80$\pm$0.11& 7.72$\pm$0.05 	& 8.00$\pm$0.08 \\
S$^{+}$	 &6.32$\pm$0.05 & 6.49$\pm$0.06 & 6.17$\pm$0.05 & 6.29$\pm$0.06& 5.09$\pm$0.10  & 5.18$\pm$0.10 \\
S$^{++}$ & 6.84$\pm$0.06& 7.22$\pm$0.10 & 6.79$\pm$0.06 & 7.09$\pm$0.10& 6.83$\pm$0.04  & 7.11$\pm$0.09 \\
Cl$^{+}$ & 4.77$\pm$0.05& 4.91$\pm$0.07 & 4.75$\pm$0.05 & 4.85$\pm$0.07& 3.46$\pm$0.07  & 3.54$\pm$0.07 \\
Cl$^{++}$& 5.04$\pm$0.06& 5.36$\pm$0.08 & 4.99$\pm$0.07 & 5.21$\pm$0.08& 5.06$\pm$0.05  & 5.30$\pm$0.08 \\
Cl$^{3+}$& ---   	& ---		& --- 		&  ---         & 3.86$\pm$0.04  & 4.06$\pm$0.07 \\
Ar$^{++}$& 6.25$\pm$0.05& 6.53$\pm$0.08 & 6.17$\pm$0.06 & 6.36$\pm$0.08& 6.35$\pm$0.04  & 6.56$\pm$0.07 \\
Ar$^{3+}$& 3.89$\pm$0.22& 4.23$\pm$0.23 & 4.01$\pm$0.18 & 4.24$\pm$0.19& 4.78$\pm$0.06  & 5.04$\pm$0.08 \\
Fe$^{+}$ & 4.62:	& 4.78:		& 4.51:		& 4.62:	       & 4.04:  	& 4.13:	\\
Fe$^{++}$& 5.07$\pm$0.04& 5.41$\pm$0.08 & 5.23$\pm$0.10 & 5.47$\pm$0.12& 5.24$\pm$0.06  & 5.50$\pm$0.09 \\
\noalign{\vskip3pt} \noalign{\hrule} \noalign{\vskip3pt}
\end{tabular}
\begin{description}
\item[$^{\rm a}$] In units of 12+log(X$^m$/H$^+$).
\end{description}
\end{minipage}
\end{table*}

Many {\ffeii} lines have been identified in our spectra, but are severely affected 
by fluorescence effects \citep{rodriguez99, verneretal00}. The {\ffeii} $\lambda8617$ ${\rm \AA}$ line is 
almost insensitive to fluorescence effects, but 
unfortunately it is in one of our narrow observational gaps.
We have also measured {\ffeii} $\lambda7155$, a line which does not seem to be affected by fluorescence effects 
\citep{rodriguez96}. We have derived the Fe$^+$ abundance from this line assuming 
that $I(\lambda7155)$/$I(\lambda8616)$$\sim$1 \citep{rodriguez96} and using the calculations of 
\citet{bautistapradhan96}. We find Fe$^+$/H$^+$ $\sim$ 4.2$\times$10$^{-8}$, 3.2$\times$10$^{-8}$ and 
1.1$\times$10$^{-8}$ for M16, M20 and NGC~3603, respectively.  In NGC~3603, the Fe$^+$ abundance 
is much lower than that of Fe$^{++}$ (see Table~\ref{celabun}). Therefore, in what follows the Fe$^{+}$ 
abundance will be considered negligible for this object. 

The calculations for Fe$^{++}$ have been done with a 34 level 
model-atom that uses the collision strengths of \citet{zhang96} and the transition probabilities of 
\citet{quinet96}. We have used 5 {\ffeiii} lines for M16, 6 for M20 and 5 for NGC~3603, 
that do not seem to be affected by line-blending. 
The Fe$^{++}$ abundances are also included in Table ~\ref{celabun}.

\subsection{Ionic Abundances from Recombination Lines}\label{rls}

\setcounter{table}{6}
\begin{table*}
\centering
\begin{minipage}{200mm}
\caption{C$^{++}$/H$^+$ abundance ratio from {\cii} lines}
\label{ciirl}
\begin{tabular}{ccccccccccc}
\noalign{\hrule} \noalign{\vskip3pt}
& & \mc{3}{c}{M16}& \mc{3}{c}{M20} & \mc{3}{c}{NGC 3603} \\
\noalign{\hrule} \noalign{\vskip3pt}
& & $I$($\lambda$)/$I$(H$\beta$) & \multicolumn{2}{c}{C$^{++}$/H$^+$ ($\times$10$^{-5}$)}& $I$($\lambda$)/$I$(H$\beta$) & \multicolumn{2}{c}{C$^{++}$/H$^+$ ($\times$10$^{-5}$)}& $I$($\lambda$)/$I$(H$\beta$) & \multicolumn{2}{c}{C$^{++}$/H$^+$ ($\times$10$^{-5}$)} \\
Mult. & $\lambda_0$ & ($\times$10$^{-2}$) & A & B & ($\times$10$^{-2}$) & A & B& ($\times$10$^{-2}$) & A & B\\
\noalign{\vskip3pt} \noalign{\hrule} \noalign{\vskip3pt}
2	& 6578.05	& 0.310$\pm$0.019	& 365$\pm$22	& 60$\pm$4 	& 0.356$\pm$0.021	& 414$\pm$25	& 68$\pm$4 	& 0.250$\pm$0.023	& 264$\pm$24	& 47$\pm$4 	\\ 
3	& 7231.12	& 0.096$\pm$0.007	& 2533$\pm$177	& 36$\pm$3 	& 0.075$\pm$0.007 	& 1971$\pm$177	& 28$\pm$3 	& 0.086$\pm$0.006	& 2300$\pm$161	& 33$\pm$2 	\\
	& 7236.19	& 0.178$\pm$0.012	& 2660$\pm$200	& 38$\pm$2 	& 0.126$\pm$0.009 	& 1884$\pm$132 	& 27$\pm$2 	& 0.179$\pm$0.011	& 2705$\pm$162	& 38$\pm$2 	\\
	& Sum		& 			& 2614$\pm$132	& 37$\pm$1 	&  			& 1915$\pm$106	& 27$\pm$1	& 			& 2558$\pm$115	& 36$\pm$1 	\\  
4	& 3918.98	& 0.138$\pm$0.018	& 2840$\pm$369	& 900$\pm$117 	& 0.094$\pm$0.019	& 1885$\pm$377	& 595$\pm$119	& ---			& ---		& --- 		\\
	& 3920.68	& 0.162$\pm$0.019	& 1660$\pm$199	& 525$\pm$63 	& 0.141$\pm$0.020	& 1420$\pm$199	& 450$\pm$63	& ---			& ---		& --- 		\\
	& Sum		& 			& 2060$\pm$175	& 650$\pm$55	& 			& 1575$\pm$176	& 500$\pm$56	& ---			& ---		& --- 		\\ 
6	& 4267.26	& 0.272$\pm$0.019	& 25$\pm$2	& {\bf 25$\pm$2}& 0.170$\pm$0.020	& 15$\pm$2	& {\bf 15$\pm$2}& 0.325$\pm$0.059	& 31$\pm$6	& {\bf 30$\pm$5} \\ 
17.02	& 9903.46	& 0.037$\pm$0.005$^{\rm a}$& 13$\pm$2	& ---		& 0.066:		& 24:		& ---		& 0.111$\pm$0.010$^{\rm b}$& 43$\pm$4	& --- 		\\ 
17.04	& 6461.95	& 0.032$\pm$0.012	& 28$\pm$10	& ---		& 0.043$\pm$0.011$^{\rm c}$& 38$\pm$10	& ---		&  --- 			& ---		& --- 		\\ \hline 
	& Adopted	& 			&\mc{2}{c}{25$\pm$2 }		& 			&\mc{2}{c}{15$\pm$2}		& 			&\mc{2}{c}{30$\pm$ 5 } \\
\noalign{\vskip3pt} \noalign{\hrule} \noalign{\vskip3pt}
\end{tabular}
\begin{description}
\item[$^{\rm a}$] Affected by atmospheric absorption bands.
\item[$^{\rm b}$] Blend with an unidentified line.
\item[$^{\rm c}$] Affected by internal reflections or charge transfer in the CCD.
\end{description}
\end{minipage}
\end{table*}

\setcounter{table}{7}
\begin{table*}
\centering
\begin{minipage}{250mm}
\caption{O$^{+}$/H$^+$ ratio from {\oi} permitted lines$^{\rm a}$}
\label{oirl}
\begin{tabular}{ccccccccccc}
\noalign{\hrule} \noalign{\vskip3pt}
& & \mc{3}{c}{M16}& \mc{3}{c}{M20} & \mc{3}{c}{NGC 3603} \\
\noalign{\hrule} \noalign{\vskip3pt}
& & $I$($\lambda$)/$I$(H$\beta$) & \mc{2}{c}{O$^{+}$/H$^+$ ($\times$10$^{-5}$)}& $I$($\lambda$)/$I$(H$\beta$) & \mc{2}{c}{O$^{+}$/H$^+$ ($\times$10$^{-5}$)}& $I$($\lambda$)/$I$(H$\beta$) & \mc{2}{c}{O$^{+}$/H$^+$ ($\times$10$^{-5}$)} \\
Mult. & $\lambda_0$ & ($\times$10$^{-2}$) & A & B & ($\times$10$^{-2}$) & A & B & ($\times$10$^{-2}$) & A & B \\
\noalign{\vskip3pt} \noalign{\hrule} \noalign{\vskip3pt}
1& 7771.94& 0.026$\pm$0.006$^{\rm b}$& 26$\pm$6/34$\pm$8& -- & 0.036$\pm$0.006 & 36$\pm$6/47$\pm$8 & -- & -- & -- & -- \\
4& 8446.48& 0.482$\pm$0.034& 1849$\pm$129/& 413$\pm$29/ & 0.362$\pm$0.018& 1383$\pm$69/& 311$\pm$16/ & 0.196$\pm$0.014& 728$\pm$51/& 171$\pm$12/ \\ 
&  &  & 2717$\pm$190& 546$\pm$38 & & 2052$\pm$103& 412$\pm$21 & & 1165$\pm$82& 2325$\pm$163 \\ \hline
& Adopted& \mc{3}{c}{30$\pm$7 } &\mc{3}{c}{42$\pm$7} &\mc{3}{c}{---} \\ 
\noalign{\vskip3pt} \noalign{\hrule} \noalign{\vskip3pt}
\end{tabular}
\begin{description}
\item[$^{\rm a}$] Recombination coefficientes by \citet{pequignotetal91}/\citet{escalantevictor92}.
\item[$^{\rm b}$] Blended with telluric emission lines.
\end{description}
\end{minipage}
\end{table*}

\setcounter{table}{8}
\begin{table*}
\centering
\begin{minipage}{175mm}
\begin{tiny}
\caption{O$^{++}$/H$^+$ ratio from {\oii} permitted lines$^{\rm a}$}
\label{oiirl}
\begin{tabular}{ccccccccccc}
\noalign{\hrule} \noalign{\vskip3pt}
& & \mc{3}{c}{M16}& \mc{3}{c}{M20} & \mc{3}{c}{NGC 3603} \\
\noalign{\hrule} \noalign{\vskip3pt}
& & $I$($\lambda$)/$I$(H$\beta$) & \mc{2}{c}{O$^{++}$/H$^+$ ($\times$10$^{-5}$)}& $I$($\lambda$)/$I$(H$\beta$) & \mc{2}{c}{O$^{++}$/H$^+$ ($\times$10$^{-5}$)}& $I$($\lambda$)/$I$(H$\beta$) & \mc{2}{c}{O$^{++}$/H$^+$ ($\times$10$^{-5}$)} \\
Mult. & $\lambda_0$ & ($\times$10$^{-2}$) & A & B & ($\times$10$^{-2}$) & A & B & ($\times$10$^{-2}$) & A & B \\
\noalign{\vskip3pt} \noalign{\hrule} \noalign{\vskip3pt}
1$^{\rm b}$& 4638.85& 0.040$\pm$0.010	& 37$\pm$9/24$\pm$6	& 35$\pm$9/23$\pm$6	& --- 			&  --- 			&  --- 			& 0.057: 		& 52:/43: 		& 50:/41: 		\\
& 4641.81	& 0.034$\pm$0.010	& 13$\pm$3/15$\pm$4	& 13$\pm$3/14$\pm$4	& 0.030$\pm$0.015	& 12$\pm$6/14$\pm$7	& 11$\pm$6/13$\pm$7 	& 0.131$\pm$0.033	& 49$\pm$12/51$\pm$13	& 47$\pm$12/49$\pm$12 	\\ 
& 4649.14	& 0.056$\pm$0.011	& 12$\pm$2/20$\pm$4	& 11$\pm$2/19$\pm$4	& 0.036$\pm$0.018	& 8$\pm$4/20$\pm$10	& 7$\pm$4/19$\pm$10 	& 0.172$\pm$0.034	& 36$\pm$7/42$\pm$8	& 35$\pm$7/41$\pm$8 	\\
& 4650.84	& 0.050$\pm$0.011	& 49$\pm$11/28$\pm$6	& 47$\pm$11/27$\pm$6	& 0.030$\pm$0.015 	& 17$\pm$9/8$\pm$4	& 16$\pm$8/7$\pm$4	& 0.086$\pm$0.029	& 84$\pm$28/65$\pm$22	& 81$\pm$27/63$\pm$21 	\\
& 4661.64	& 0.038$\pm$0.010	& 31$\pm$8/20$\pm$5	& 30$\pm$8/19$\pm$5	& 0.018$\pm$0.014	& 15$\pm$12/8$\pm$6	& 14$\pm$11/8$\pm$6 	& 0.111$\pm$0.031	& 90$\pm$25/74$\pm$21	& 87$\pm$24/72$\pm$20	\\
& 4676.24	& --- 			& --- 			& --- 			& --- 			& --- 			& --- 			& 0.040:		& 42:/44:		& 41:/43: 		\\
& 4696.36	& --- 			& --- 			& --- 			& --- 			& --- 			& --- 			& 0.037$^{\rm c}$:	& 381:/325:		& 368:/304: 		\\
& Sum		& 			& 21$\pm$2		& {\bf 20$\pm$2}	&  			& 11$\pm$5		& {\bf 10$\pm$5} 	& 			& 53$\pm$6		& {\bf 51$\pm$6} 	\\ \hline
2& 4349.43	& 			& 			& 			& 			& 			& 			& 0.148:		& 32:			& 31: 			\\ \hline
& Adopted	& 			&\mc{2}{c}{\bf 20$\pm$2 }			& 			&\mc{2}{c}{\bf 10$\pm$5 }			& 			&\mc{2}{c}{\bf 51$\pm$6 } 			\\ 
\noalign{\vskip3pt} \noalign{\hrule} \noalign{\vskip3pt}
\end{tabular}
\end{tiny}
\begin{description}
\item[$^{\rm a}$] Except in the case of M20, only lines with intensity uncertainties lower than 40 \% have been considered (see text)
\item[$^{\rm b}$] Not corrected from NLTE effects/Corrected from NLTE effects (see text).
\item[$^{\rm c}$] Probably is a misidentification
\end{description}
\end{minipage}
\end{table*}

We have measured a large number of permitted lines of heavy element ions such as {\oi}, {\oii}, 
{\ci}, {\cii}, {\sii}, {\nitroi}, {\nii}, {\ari}, {\sili}, {\silii}, and {\fei}
many of them detected for the first time in these nebulae. Unfortunately, most permitted lines are affected by fluorescence 
effects or blended with telluric emission lines making their intensities unreliable. 
Detailed discussions on the mechanism of formation of the permitted lines can be found in 
\citet[][and references therein]{estebanetal98, estebanetal04}.

\begin{figure*}
\begin{center}
\epsfig{file=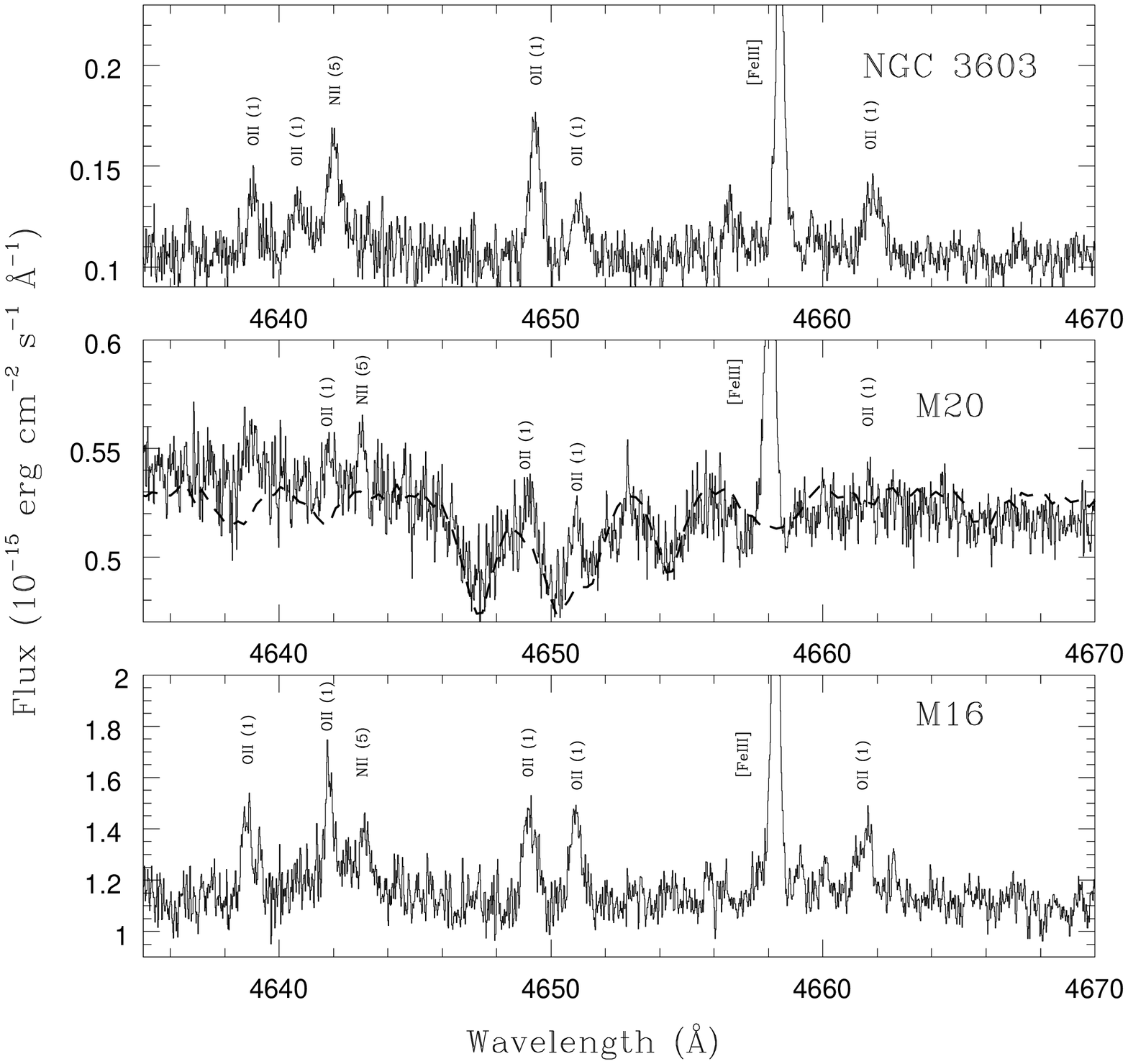, width=15. cm, clip=}
\caption{Section of the echelle spectra showing the lines of multiplet 1 of {\oii} (observed fluxes) for the three {\hii} regions. 
In the case of M20, we have 
superimposed the spectrum of HD164492 (dashed line), which is normalized to the continuum flux in the zone of {\oii} 4649 and 4650 {\rm \AA} lines. 
It can be seen that the fluxes of multiplet 1 emission lines may be measured simply integrating the line flux between the closest 
points of the local adyacent continuum, and that 
these lines are not seriously attenuated by the dust-scattered light (see text).
\label{m1oii}}
\end{center}
\end{figure*}

For the first time for these nebulae, we have been able to measure the ionic abundance ratios of 
O$^{+}$/H$^+$, O$^{++}$/H$^+$ and C$^{++}/$H$^+$ from pure 
recombination lines. 
We have computed the abundances using {\te}(Low) (for O$^{+}$/H$^+$), {\te}(High) (for O$^{++}$/H$^+$ and 
C$^{++}/$H$^+$) and {\elecd} from Table~\ref{plasma}. 
Atomic data and methodology are the same than in \citet{garciarojasetal04}. 
Although part of these ionic abundances were presented in a previous work of our group \citep{estebanetal05}, 
we give here the details of their derivation. 

Eight permitted lines of {\cii} have been measured in M16 and M20, and only five in NGC~3603. 
Lines of multiplets 6, 17.02 and 17.04 are $3d-4f$ transitions and are, in principle, excited by pure 
recombination \citep[see][]{grandi76}. Unfortunately, only multiplet 6 is usable because the lines of the other 
multiplets are affected by blending with atmospheric spectral features or CCD charge transfer effects, so we 
have adopted the C$^{++}/H$$^+$ ratio given by that multiplet (see Table~\ref{ciirl}). 

The O$^{+}$ abundance was derived from the {\oi} $\lambda$7771.94 \AA\ line, and was only reliable for M20, 
because the spectral zone of multiplet 1 is strongly affected by telluric lines. The abundance derived 
from this line is case independent and recombination is its formation mechanism because the line corresponds to a 
quintuplet transition, while the ground term of this ion is a triplet. The O$^{+}$/H$^+$ ratios are presented in Table~\ref{oirl}. 

We have measured several lines of multiplet 1 of {\oii}. As it has been pointed out by \citet{tsamisetal03} and 
\citet{ruizetal03}, the upper levels of the transitions of multiplet 1 of {\oii} are not in LTE for densities 
$n_e$$<$10000 cm$^{-3}$, and the abundances derived from each individual line could differ by factors as large as 4. 
We have applied the NLTE 
corrections estimated by \citet{apeimbertetal05} to our data and the abundances obtained from the individual 
lines are in good agreement and also agree with the abundance derived using the sum of all the lines of the 
multiplet, which is not affected by NLTE effects. 
On the other hand, \citet{tsamisetal03} pointed out that, in the presence of absorption line features in the multiplet 1 spectral 
range, the emission lines could be attenuated. This effect can be very important in extragalactic objects, and it can only be 
corrected if the stars are resolved, or if synthetic spectra are available. In our case, our high resolution spectra shows, 
when compared with the spectrum of HD164492 (kindly provided by S. Sim\'on-D\'{\i}az), the main ionizing source of M20, 
that the continuum does not affect the 
measurement of multiplet 1 {\oii} emission line fluxes, in spite of the large fraction of dust-scattered light of the nebular continuum 
(see \S~\ref{contm20}). Indeed, the situation may be quite different in the case of low spectral resolution observations, 
as was the case of \citet{tsamisetal03} for 30 Dor and LMC N11B.
The O$^{++}$/H$^+$ ratios for the three nebulae are presented in Table~\ref{oiirl}. 

\section{Total abundances}\label{totabun}

We have adopted a set of ionisation correction factors (ICF) to correct for the unseen ionisation 
stages and then derive the total gaseous abundances of the chemical elements we have studied. We have 
adopted essentially the ICF scheme used by \citet{garciarojasetal05} for all the elements, but we 
will discuss some special cases.

The absence of {\heii} lines in our spectra indicates that He$^{++}$/H$^+$ is 
negligible. However, the total helium abundance has to be corrected for the presence 
of neutral helium. Based on the ICF(He$^0$) given by \citet{peimbertetal92} and with our data, the ICF(He$^0$) 
amounts to 1.18$\pm$0.05, 1.16$\pm$0.05 and 1.007$\pm$0.002 for {\ts}$>$0.00, for M16, M20 
and NGC~3603, respectively.

For all the nebulae, we have derived the O/H ratio from CELs, from the combination of O$^{++}$/H$^+$ ratio 
from RLs and O$^{+}$/H$^+$ ratio from CELs and the assumed {\ts} (for M16 and NGC~3603) and, for the first time for M20, 
from pure recombination lines of O$^+$ and O$^{++}$. 
In Table \ref{total} we show the total abundances obtained in our nebulae
for {\ts}=0.00 and {\ts}$>$0.00. 

\setcounter{table}{9}
\begin{table*}
\begin{minipage}{150mm}
\centering \caption{Total Gaseous Abundances.}
\label{total}
\begin{tabular}{lcccccc}
\noalign{\hrule} \noalign{\vskip3pt}
& \mc{2}{c}{M16}& \mc{2}{c}{M20} & \mc{2}{c}{NGC 3603} \\
\noalign{\vskip3pt} \noalign{\hrule} \noalign{\vskip3pt}
Element & {\ts}=0.000 & {\ts}=0.039$\pm$0.006 & {\ts}=0.000 & {\ts}=0.029$\pm$0.007 &  {\ts}=0.000 & {\ts}=0.040$\pm$0.008 \\
\noalign{\vskip3pt} \noalign{\hrule} \noalign{\vskip3pt}
He	   & 11.01$\pm$0.02 	& 10.97$\pm$0.02& 10.95$\pm$0.06& 10.92$\pm$0.06& 10.99$\pm$0.01& 10.99$\pm$0.01\\
C	   & 8.76$\pm$0.06 	& 8.76$\pm$0.06 & 8.69$\pm$0.08 & 8.69$\pm$0.08	& 8.51$\pm$0.07 & 8.51$\pm$0.07 \\
N	   & 7.84$\pm$0.06 	& 8.07$\pm$0.12 & 7.67$\pm$0.05 & 7.83$\pm$0.07 & 7.62$\pm$0.13 & 7.89$\pm$0.14 \\
O	   & 8.56$\pm$0.07	& 8.78$\pm$0.07 & 8.53$\pm$0.06	& 8.67$\pm$0.07 & 8.46$\pm$0.05	& 8.71$\pm$0.07 \\
O$^{\rm a}$& 8.81$\pm$0.07	& 8.81$\pm$0.07 & 8.71$\pm$0.07	& 8.71$\pm$0.07 & 8.72$\pm$0.05	& 8.72$\pm$0.05 \\
Ne	   & 7.86$\pm$0.15 	& 8.08$\pm$0.17 & 7.83$\pm$0.16 & 7.97$\pm$0.18 & 7.76$\pm$0.08 & 8.03$\pm$0.11 \\
S	   & 6.96$\pm$0.05 	& 7.29$\pm$0.08 & 6.88$\pm$0.05 & 7.12$\pm$0.09 & 7.03$\pm$0.05 & 7.36$\pm$0.08 \\
Cl$^{\rm b}$& 5.23$\pm$0.04	& 5.49$\pm$0.07	& 5.19$\pm$0.05	& 5.37$\pm$0.06 & 5.09$\pm$0.05	& 5.33$\pm$0.07 \\
Ar	   & 6.70$\pm$0.07 	& 6.84$\pm$0.08 & 6.65$\pm$0.06 & 6.70$\pm$0.11 & 6.37$\pm$0.15 & 6.58$\pm$0.17 \\
Fe$^{\rm c}$ & 5.17$\pm$0.11	& 5.53$\pm$0.13 & 5.31$\pm$0.13 & 5.56$\pm$0.15 & 6.14$\pm$0.16/5.74$\pm$0.10 & 6.53$\pm$0.19/6.05$\pm$0.10 \\
Fe$^{\rm d}$ & 5.20$\pm$0.06	& 5.51$\pm$0.07 & 5.31$\pm$0.09	& 5.52$\pm$0.10	& 5.27$\pm$0.06	& 5.52$\pm$0.09 \\
\noalign{\vskip3pt} \noalign{\hrule} \noalign{\vskip3pt}
\end{tabular}
\begin{description}
\item[$^{\rm a}$] For M20, O$^{+}$/H$^+$ and O$^{++}$/H$^+$ from RLs. For M16 and NGC~3603, O$^{++}$/H$^+$ from RLs and O$^+$/H$^+$ from CELs and t$^2$.
\item[$^{\rm b}$] For NGC~3603, from Cl$^{+}$/H$^+$+Cl$^{++}$/H$^+$+Cl$^{3+}$/H$^+$. For M16 and M20, from Cl$^{+}$/H$^+$+Cl$^{++}$/H$^+$. 
\item[$^{\rm c}$] ICF from equation 2 for M16 and M20; ICF from equation2/equation 3 for NGC~3603.
\item[$^{\rm d}$] Fe/H = Fe$^+$/H$^+$ + Fe$^{++}$/H$^+$.
\end{description}
\end{minipage}
\end{table*}

For neon, we have used the ICF proposed by \citet{peimbertcostero69}:

\begin{equation}
\frac{N(\rm Ne)}{N(\rm H)} =
            \left( \frac{N({\rm O^+})+N(\rm O^{++})}{N(\rm O^{++})} \right)
            \frac{N(\rm Ne^{++})}{N(\rm H^+)}.
\end{equation}

Nevertheless this ICF underestimates the Ne/H
abundance for nebulae of low degree of ionisation because a considerable
fraction of Ne$^+$ coexists with O$^{++}$ 
\citep[see][]{torrespeimbertpeimbert77,peimbertetal92}. This is the case for
M16 and M20.  Based on the O$^+$/O ratio, the data and the prescriptions by \citet{torrespeimbertpeimbert77},
we estimate that the ICF(Ne)'s
should be about 0.14  $\pm$  0.1 dex for M16 and 0.42  $\pm$  0.1 dex for M20 higher than those
provided by  equation  (1) for {\ts}=0.00. From these ICF(Ne) 's we derive an Ne/O ratio of about
0.2 for both regions. This ratio is  in excellent agreement with the Ne/O ratios derived for M17
by \citet{peimbertetal92}, and by us for NGC 3603, where most of the O
and Ne are twice ionized and the ICF(Ne) is very small.
Given the high ionisation degree for NGC~3603, equation 1 is a good approximation to the fraction of Ne$^{+}$ in 
this nebula.

We have measured lines of two ionisation stages of chlorine in M16 and M20: 
Cl$^{+}$ and Cl$^{++}$. The Cl abundance has been assumed to be equal to 
the sum of these ionic abundances without taking into account the Cl$^{3+}$ 
fraction. This assumption seems reasonable taking into account the small 
Cl$^{3+}$/Cl$^{++}$ ratio found for M17 \citep[$\sim$0.03, see][]{estebanetal99a}, 
for the Orion nebula \citep[$\sim$0.04, see][]{estebanetal04}, for NGC~3576 
\citep[$\sim$0.02, see][]{garciarojasetal04}, and for NGC~3603 ($\sim$0.06, this work) 
and the lower ionisation degree of M16 and 
M20 with respect to those nebulae. In NGC~3603 we have detected three 
ionisation stages of chlorine and the total abundance includes the sum of 
the Cl$^{+}$, Cl$^{++}$ and Cl$^{3+}$ abundances. 
Using the ICF by \citet{peimberttorrespeimbert77} 
to correct for the the presence of Cl$^{3+}$, we have obtained abundances 0.13 
and 0.07 dex higher for M16 and M20, and 0.02 dex lower for NGC~3603, showing that 
this ICF scheme is a good approximation when {\fcliv} lines are not detected 
in the spectrum of {\hii} regions with high degree of ionisation, however, it 
overestimates the contribution of Cl$^{3+}$ for the low ionisation regime.

We have measured lines of two stages of ionisation of iron: Fe$^+$ and Fe$^{++}$. As we have commented 
in \S~\ref{cels}, the Fe$^+$ abundance is somewhat uncertain, so we have used the ICF scheme by 
\citet{rodriguezrubin05} (based on photoionization models) to derive the total Fe/H ratio from the Fe$^{++}$ 
abundance, which is given by:

\begin{equation}
\frac{N({\rm Fe})}{N({\rm H})} = 0.9\left[\frac{N({\rm O}^{+})}{N({\rm O}^{++})}\right]^{0.08}\times\frac{N({\rm Fe})^{++}}{N({\rm O})^{+}}\times\frac{N({\rm O})}{N({\rm H})},
\end{equation}

In the case of high ionisation degree nebulae, \citet{rodriguezrubin05} 
used a further relation based on an observational fit, which is given by:

\begin{equation}
\frac{N({\rm Fe})}{N({\rm H})} = 1.1\left[\frac{N({\rm O}^{+})}{N({\rm O}^{++})}\right]^{0.58}\times\frac{N({\rm Fe})^{++}}{N({\rm O})^{+}}\times\frac{N({\rm O})}{N({\rm H})};
\end{equation}

This last relation has been applied to obtain the Fe/H ratio of NGC~3603. The discrepancy observed between the Fe abundance 
obtained making use of equation (2) or (3) for high ionisation degree nebulae (e.g. see NGC~3603 Fe abundance on Table~\ref{total}) 
has been extensively discussed by \citet{rodriguezrubin05}. 
From Table~\ref{total} it is clear that the 
sum of Fe$^{+}$ and Fe$^{++}$ abundances for M16 and M20 respectively 
are almost coincident with those derived using an ICF. In fact, for these regions,  
it is not expected a large contribution of Fe$^{3+}$ to the total abundance, due to their low ionisation degree. 
This is not true for objects with high ionisation degree as NGC~3603, for which the fraction of Fe$^{+3}$ is expected 
to be large; this is reflected in the large difference between the two values of Fe/H ratio given in Table~\ref{total}.  
It is obvious that the sum of Fe$^{+}$ and Fe$^{++}$ abundances is not applicable for this object, we must rely on the 
results obtained assuming an ICF.

\section{Deuterium Balmer Lines in M16 and M20}\label{deut}

\begin{figure}
\begin{center}
\epsfig{file=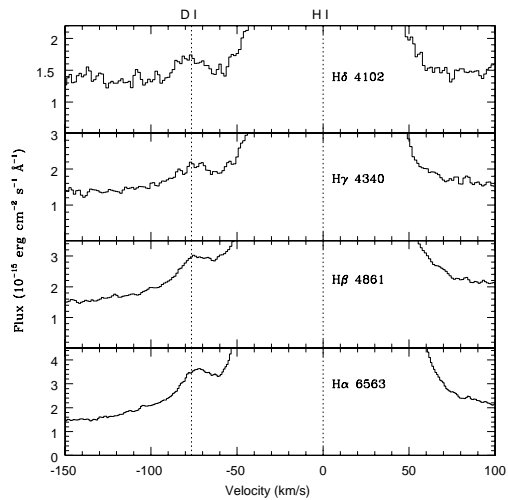, width=7.5 cm, clip=}
\caption{Wings of {\ha} to H$\delta$ in M16. The lines are centred at 0 km s$^{-1}$ 
velocity. The dotted line of the left correspond to the average wavelength adopted 
for the {\di} lines.
\label{deum16}}
\end{center}
\end{figure}

\begin{figure}
\begin{center}
\epsfig{file=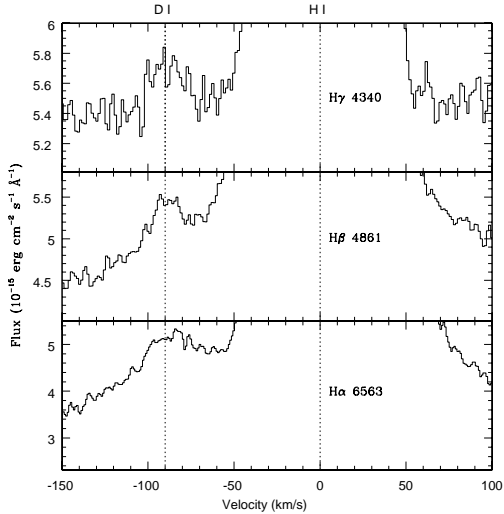, width=7.5 cm, clip=}
\caption{Same as figure 4 for M20.
\label{deum20}}
\end{center}
\end{figure}

\citet{hebrardetal00b} reported the detection of deuterium Balmer lines 
in the spectrum of M16 and M20. These authors 
detected from D$\alpha$ to D$\gamma$ in M16 and only D$\alpha$ in M20. 
We have detected several deuterium Balmer lines in M16 --from D$\alpha$ to D$\delta$-- and in M20 
--from D$\alpha$ to D$\gamma$--; these lines appear 
as very weak emission features in the blue wings of the corresponding {\hi} Balmer lines  
(see figures~\ref{deum16} and \ref{deum20}). The apparent shifts in radial velocity of 
these lines with respect to the hydrogen ones are $-76.5$ km s$^{-1}$ for M16 and $-90.8$ km s$^{-1}$ 
for M20, which are roughly consistent with the expected isotopic shift of deuterium, $-81.6$ km s$^{-1}$.

\setcounter{table}{10}
\begin{table}
\begin{minipage}{75mm}
\centering \caption{Deuterium Balmer line properties in M16 and M20.}
\label{deuchar}
\begin{tabular}{ccccc}
\noalign{\hrule} \noalign{\vskip3pt}
Line & {\di} Isotopic & FWHM {\di} & FWHM {\hi} &{\di}/{\hi} ratio \\
     &  shift (km s$^{-1}$) & (km s$^{-1}$)   & (km s$^{-1}$) & ($\times$10$^{-4}$)  \\
\noalign{\vskip3pt} \noalign{\hrule} \noalign{\vskip3pt}
M16 & & & & \\
\noalign{\vskip3pt} \noalign{\hrule} \noalign{\vskip3pt}
$\alpha$ & $-75.4$ & $<$10: & 24 & 1.0 $\pm$ 0.3  \\
$\beta$  & $-76.5$ & $<$10: & 24 & 3.2 $\pm$ 0.9 \\
$\gamma$ & $-76.7$ & $<$10: & 24 & 6.4 $\pm$ 2.4 \\
$\delta$ & $-77.5$ & $<$10: & 24 & 8.2:       \\
\noalign{\vskip3pt} \noalign{\hrule} \noalign{\vskip3pt}
M20 & & & & \\
\noalign{\vskip3pt} \noalign{\hrule} \noalign{\vskip3pt}
$\alpha$ & $-92.7$ & $<$10:  & 25  & 2.0 $\pm$ 0.4 \\
$\beta$  & $-90.7$ & $<$10:  & 20  & 3.9 $\pm$ 1.5 \\
$\gamma$ & $-89.1$ & $<$10:  & 21  & 4.7:  \\
\noalign{\vskip3pt} \noalign{\hrule} \noalign{\vskip3pt}
\end{tabular}
\end{minipage}
\end{table}

We have excluded the possibility that these weak features are high velocity components of hydrogen following the same criteria 
as \citet{garciarojasetal05} for the case of S~311. The first criterion is the absence of similar high-velocity 
components associated to bright lines of other ions. The second is that the full width at half maximum (FWHM) of the deuterium 
lines is narrower than 10 km s$^{-1}$ in all cases, much narrower than the {\hi} Balmer lines (see Figures~\ref{deum16} 
and \ref{deum20}), which have FWHM between 20 and 25 km s$^{-1}$. This fact supports the idea that deuterium lines arise from 
much colder material, probably from the photon dominated region \citep{hebrardetal00b}.

In order to strengthen the conclusions about the nature of the emission of 
{\di} Balmer lines we have compared the Balmer decrements of the hydrogen and deuterium lines observed 
in our spectra with the standard fluorescence models by \citet[][see their Figure 13]{odelletal01} 
for the Orion nebula, finding that our observations follow closely this model, indicating that fluorescence should be the 
main excitation mechanism of the {\di} lines observed in M16 and M20 (see Table~\ref{deuchar}). 

\section{High-Velocity Components in NGC~3603}\label{veloc}

\begin{figure}
\begin{center}
\epsfig{file=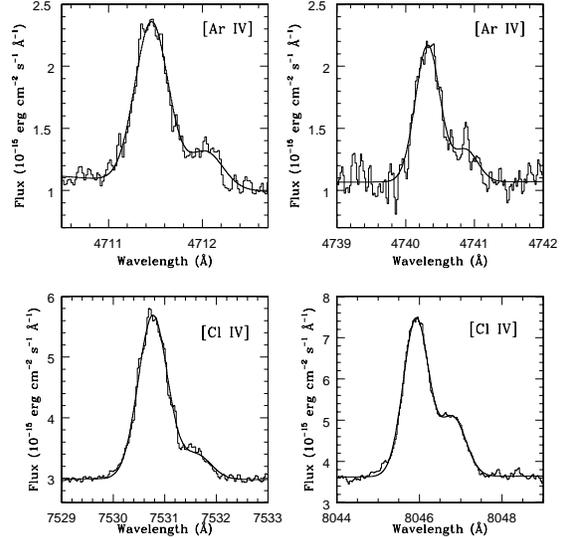, width=7.5 cm, clip=}
\caption{Redshifted components in the wings of {\fariv} and {\fcliv} lines in NGC~3603.
\label{redcomp}}
\end{center}
\end{figure}

\begin{figure}
\begin{center}
\epsfig{file=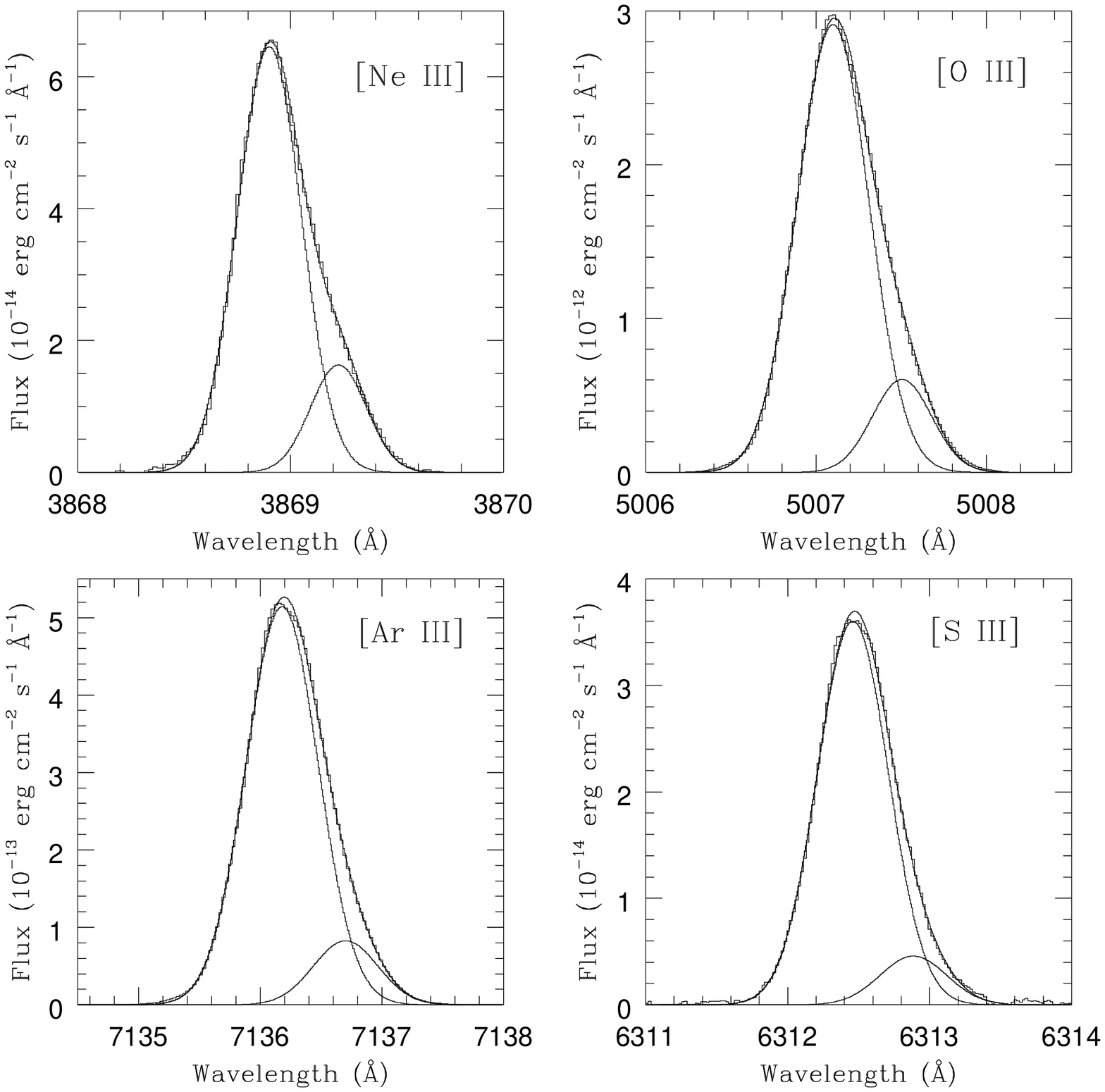, width=7.5 cm, clip=}
\epsfig{file=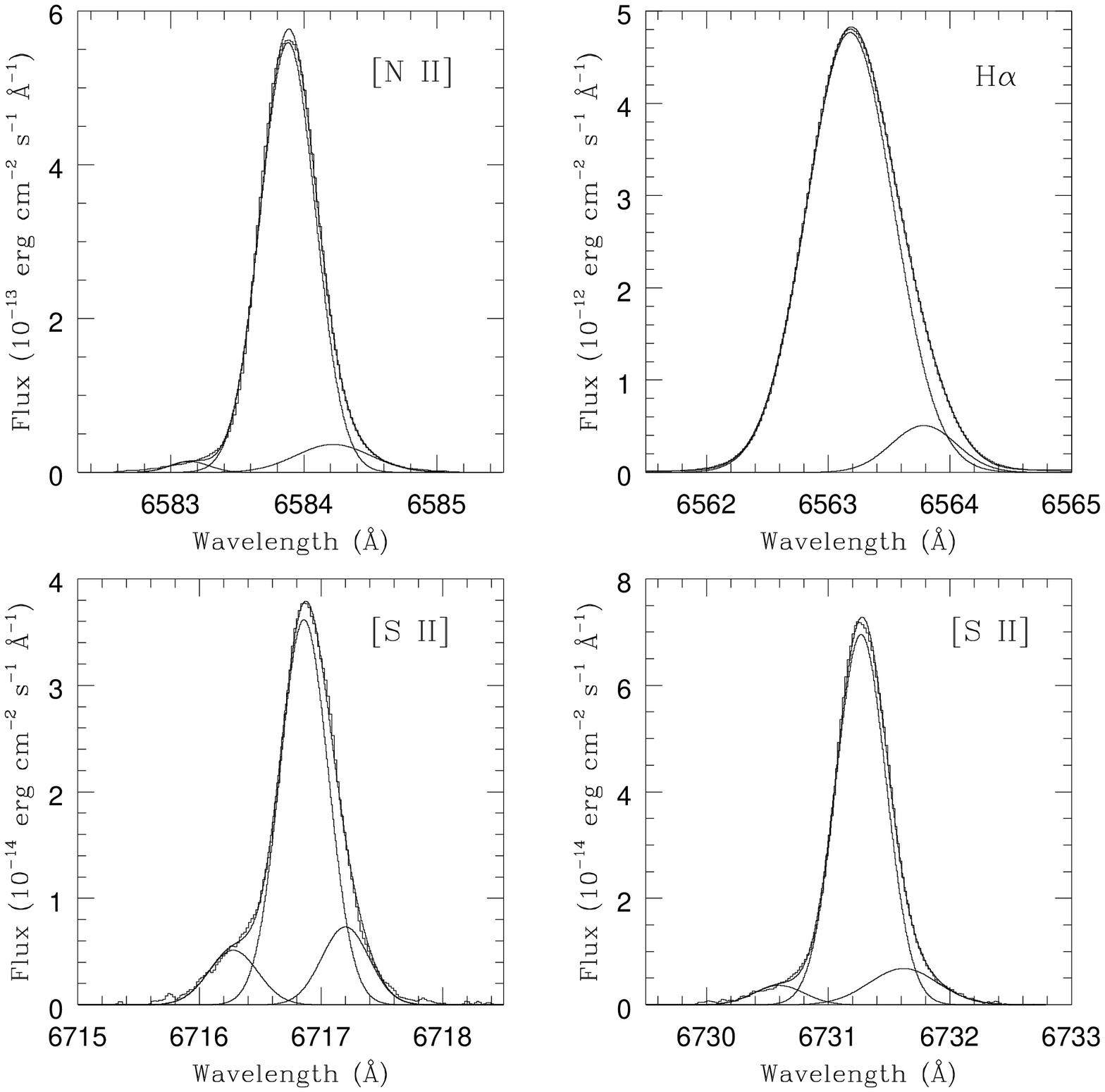, width=7.5 cm, clip=}
\caption{Components in the wings of {\fneiii}, {\foiii}, {\fariii}, {\fsiii}, {\fnii}, {\fsii} 
lines and H$\alpha$ in NGC~3603. Ionisation Potential of the ions that produce the lines decreases from top to bottom and from left to right.
\label{redcomp2}}
\end{center}
\end{figure}

We have detected weak emission features in the red wing of the highest ionisation potential lines in the spectrum of NGC~3603:  
those of {\fariv} --40.74 eV-- and {\fcliv} --39.61 eV-- (see Figure~\ref{redcomp}). 
These features are redshifted $\sim$ 36 km s$^{-1}$ (argon), and 
$\sim$33 km s$^{-1}$ (chlorine) with respect to the {\fariv} and {\fcliv} lines. The FWHM of these presumed high-velocity 
components is similar ($\sim$18 km s$^{-1}$) to that of the main component. The redshifted component is also detectable in 
the line profiles of other lines (see Figure~\ref{redcomp2}), but it is much less evident, with a contrast that decreases as the ionisation potential 
of the ion that produces the line decreases. This indicates that the redshifted component is composed by a gas with a higher 
ionisation degree than the main component. An additional fainter and blue-shifted component seems to be present in the lines 
of the low ionisation potential ions, as {\fnii} and {\fsii}. 

\citet{clayton90} obtained high spectral resolution {\foiii} profiles at  different zones of NGC~3603 finding a clear red 
component in the {\foiii} 5007 
${\rm \AA}$ line along most of the spatial extension of the nebula. At or very near our slit position, he detects that this 
second component shows a velocity shift between 30 and 40 km s$^{-1}$ with respect to the main one, values consistent with 
our reported velocity separations. The gas motions in NGC~3603 are rather complex and can be interpreted as the product of 
different expanding structures with velocities up to 100 km s$^{-1}$. 

\section{Discussion}\label{discus}

In order to improve the clarity of this paper, we have divided the discussion in 
different subsections, one devoted to M16 and M20, and another one devoted to the discussion 
of NGC~3603, whose peculiarities and its status as a Galactic example of giant {\hii} region make it a 
specially interesting object.

\subsection{M20 and M16}\label{m20m16discus}

\subsubsection{M20 Continuum determinations}\label{contm20}

\begin{figure}
\begin{center}
\epsfig{file=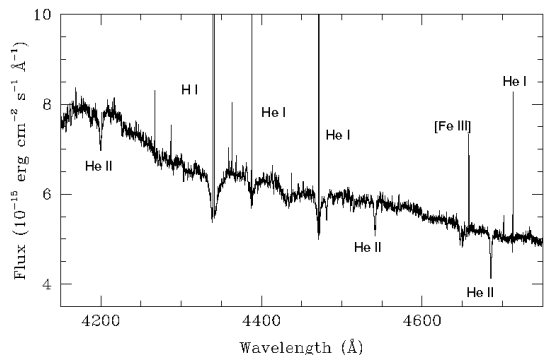, width=7.5 cm, clip=}
\caption{Section of the echelle spectrum showing the absorption $\lambda$$\lambda$4200 ${\rm \AA}$, 
4542 ${\rm \AA}$ and 4686 ${\rm \AA}$ He$^{+}$ lines.
\label{heliumii}}
\end{center}
\end{figure}

In the case of M20, our slit position is located in a bright zone very near its ionising star HD164492 (17$\arcsec$ 
north and 10$\arcsec$ east). This bright zone is just at the border of one of the dust lanes that crosses the nebula. 
Therefore, it is not strange that the stellar scattered light contribution in this spectra is specially high. This can 
be noted in the absorption features present in Figure~\ref{m1oii} and in Figure~\ref{heliumii}, where we detect the 
stellar {\heii} absorptions at $\lambda\lambda$4200, 4542 and 4686 \AA\ as well as absorption in the {\hi} Balmer lines. 

\setcounter{table}{11}
\begin{table}
\begin{minipage}{75mm}
\centering \caption{M20 Continuum determinations$^{\rm a}$}
\label{cont}
\begin{tabular}{cccc}
\noalign{\hrule} \noalign{\vskip3pt}
$\lambda$(\AA ) &\multicolumn{3}{c}{log ($j$($\lambda$)/$j$({\hb}))} \\
\noalign{\vskip3pt} \noalign{\hrule} \noalign{\vskip3pt}
& Atomic & Observed & Scattered light \\
\noalign{\vskip3pt} \noalign{\hrule} \noalign{\vskip3pt}
3640 & --2.179 &  --1.528 $\pm$ 0.007 &  --1.638 $\pm$ 0.008 \\
3670 & --3.094 &  --1.634 $\pm$ 0.004 &  --1.649 $\pm$ 0.004 \\
4110 & --3.246 &  --1.733 $\pm$ 0.003 &  --1.746 $\pm$ 0.003 \\
4350 & --3.281 &  --1.818 $\pm$ 0.004 &  --1.833 $\pm$ 0.004 \\
4850 & --3.316 &  --2.007 $\pm$ 0.004 &  --2.029 $\pm$ 0.004 \\
6570 & --3.346 &  --2.370 $\pm$ 0.006 &  --2.419 $\pm$ 0.008 \\
8175 & --3.332 &  --2.674 $\pm$ 0.002 &  --2.783 $\pm$ 0.003 \\
8260 & --4.001 &  --2.678 $\pm$ 0.008 &  --2.957 $\pm$ 0.015 \\
\noalign{\vskip3pt} \noalign{\hrule} \noalign{\vskip3pt}
\end{tabular}
\begin{description}
\item[$^{\rm a}$] in units of (${\rm \AA}^{-1}$)
\end{description}
\end{minipage}
\end{table}

In Table~\ref{cont} we show the observed and the expected nebular continua as well as the estimated scattered light 
contribution for different wavelengths. The nebular atomic continuum is the sum of the continua produced by {\hi} and {\hei} and 
it has been derived for the physical conditions and the He/H ratio computed for M20. From the table, it is evident that 
the scattered light is the main contribution to the observed continuum. As expected for normal dust properties, we have 
also found that the amount of stellar scattered light increases systematically towards bluer wavelengths (see~Table\ref{cont}). 
Using the observed equivalent widths of the {\heii} absorption lines and those detected in the spectrum of HD164492\footnote{This 
spectrum was obtained with the Intermediate Dispersion Spectrograph (IDS) attached to the INT 2.5m Telescope of the Roque de los
Muchachos observatory in La Palma, Spain. The results used here were kindly provided by Sergio Sim\'on-D\'{\i}az 
(private communication)}
it is possible to estimate the fraction of dust scattered light, 
using the expression used by \citet{sanchezpeimbert91}:

\begin{equation}
\frac{i_d(\lambda)}{i} = 
\frac{\sum{EW(M20)}}{\sum{EW({\rm HD}164492)}},
\end{equation}

where $i_d$($\lambda$)/$i$ is the fraction of dust scattered light in M20 with respect to the total emission of HD164492 at a 
given wavelength range. 
From this expression we have obtained $i_d$/$i$ = 0.54 $\pm$ 0.09; on the other hand, the fraction of the continuum scattered light with 
respect to the observed continuum in the 4200 to 4850 ${\rm \AA}$ range in M20 
amounts to 0.96 $\pm$ 0.01. \citet{odelletal66} reported that the effective dust-to-gas
ratio is 5 times higher in M20 than in other gaseous nebulae or the interstellar medium; also, \citet{robledorella02} reported 
that the nebular continuum of M20 is strongly dominated by dust-scattered light. These results 
agree with the high fraction of dust scattered light observed in the continuum of M20. 
This high fraction of dust scattered light could be due to the closeness of our slit position to HD164492, moreover, 
other nearby stars could be contributing to the observed continuum.

\subsubsection{Comparison with other abundance determinations}

As we have stated in section \ref{cels}, we have derived our ionic abundances from CELs making use of all the individual 
lines (which are not blended with other lines) of each ion observed. All the individual ionic abundances are consistent, 
within the errors, with the adopted weighted mean, except for the case of S$^+$, for which we have obtained larger differences 
because we have used {\te}{(low)} instead of {\te}({\fsii}) to derive S$^+$ abundances. In fact, the largest differences 
are between abundances derived from $\lambda$$\lambda$ 6717, 6731 ${\rm \AA}$ and from $\lambda$$\lambda$ 10329, 10336 and 
10371${\rm \AA}$, (over 0.15 dex in the case of M16), but these last lines have less weight in the final adopted value.

Previous abundance determinations in M20 and M16 are those of \citet{hawley78} 
and \citet{rodriguez98, rodriguez99b}. All of them are based on the analysis of CELs. Although the slit positions studied in 
these works are different, we have compared their results with ours. For the sake of consistency we have re-computed 
the abundances given by those authors using the same set of atomic data and ICF scheme than in this paper. Moreover, 
taking into account that the previous works obtained abundances in several slit positions across the nebula, we have 
taken average values for the comparison. 

In general there is a good agreement between our O, N and S abundances obtained from CELs and those obtained by 
\citet{hawley78} within the errors. Departures from our values are smaller than 0.1 dex for N, and the O abundance is 
almost coincident in the case of M20 and 0.09 dex lower in the case of M16. 
Taking into account that Hawley highlighted the difficulties in 
the measurement of some lines (those of {\fnii} and {\fsiii}) in his spectra, the agreement with our results is remarkable. 

We have followed the same methodology to compare with the results of \citet{rodriguez98, rodriguez99b}. 
For M20, N and S show differences of 0.06 dex and 0.09 dex respectively, 
but the O abundance shows a larger discrepancy (0.21 dex), mainly due to differences in the O$^+$/H$^+$ ratio. This 
difference can be explained because 
Rodr\'{\i}guez determined the O$^+$/H$^+$ ratio from the {\foii} 7320+7330 ${\rm \AA}$  lines, which were severely 
affected by sky telluric lines \citep{rodriguez98}. For M16 there is a good agreement between our results 
and Rodr\'{\i}guez ones, specially for N and S abundances, which are almost coincident with our values. 
The discrepancy in the O abundance is, in this case, of 0.11 dex, probably due to the same reason pointed out above.

\subsection{NGC~3603}\label{ngc3603discus}

As it was commented in the introduction, NGC~3603 is the only Galactic giant 
{\hii} region that can be observed in the visual. \citet{melnicketal89} derived an O abundance 
for NGC~3603 of 12 + log(O/H) = 8.39$\pm$0.41. In spite of its extremely large uncertainty, 
this value is in good agreement with our derived O/H abundance ratio from CELs. 
On the other hand, \citet{tapiaetal01} published the most complete set of abundances in the literature 
for NGC~3603 until now; their O/H ratio is only 0.06 dex higher than ours, and the large differences 
in the other abundance ratios are probably due to their large line intensity uncertainties and the different set of 
ICFs used. We have found a similar behaviour comparing our abundances based on CELs with those of \citet{girardietal97}.

For NGC~3603, we have available O, N, Ne and S abundance determinations based 
on far-infrared fine-structure line observations \citep{simpsonetal95}. Since the emissivity of these lines is 
essentially independent of the nebular thermal structure --due to their low excitation energies, it is interesting to 
compare abundances derived from these lines with those derived from recombination lines or from optical CELs 
assuming a {\ts}. In principle, assuming the temperature fluctuations paradigm, and that there are no 
large density fluctuations, all these determinations might be similar. In Table~\ref{comparison} 
we compare our derived total and ionic abundances for NGC~3603 (this work) 
with those obtained by \citet{simpsonetal95}. In spite of the high uncertainty of the abundances derived by 
\citet{simpsonetal95} (due to the uncertainty in the radio flux and aperture effects), values derived from IR CELs 
and those derived from optical 
CELs assuming a {\ts} $>$ 0 are similar, a fact that seems to support the presence of temperature fluctuations in the nebula. 
However, there are other examples where this is not clear; from a similar comparison between optical data and Simpson's 
IR data, other results have been found: NGC~3576 \citep{tsamisetal03, garciarojasetal05} and Orion nebula \citep{tsamisetal03, 
estebanetal04} show IR abundances which are intermediate between {\ts}=0.00 and {\ts} $>$ 0, and M17 \citep{tsamisetal03, 
estebanetal99a} and 30 Doradus \citep{apeimbert03,tsamisetal03} show IR abundances which are rather similar to those derived 
from optical CELs and {\ts}=0.00.
Moreover, the comparison of our slit optical spectroscopy and the IR data of \citet{simpsonetal95} shows a further 
complication, at least for the ionic abundances. The aperture used in both kinds of observations cover a very different 
area of the nebula, which is much larger in the IR spectroscopy. Changes in the mean ionisation degree of the area 
covered in optical and IR observations may produce natural differences in the ionic abundances not related to the 
presence or absence of a temperature structure. In contrast, this effect should not affect the total abundances. 
In this sense, we think important to clarify the conclusion drawn by \citet{tsamisetal03} about this issue. Those 
authors make the comparison between the O$^{++}$/H$^{+}$ abundances derived from optical and IR CELs for a sample 
of {\hii} regions, and conclude that temperature fluctuations might be ruled out as the cause of the discrepancy 
found between O$^{++}$/H$^{+}$ abundances derived from optical RL and CELs. From our apparently positive results 
for NGC~3603 and taking into account the aperture consideration, that conclusion seems to be far from conclusive. 
It is clear that further IR and optical observations taken in the same zones and with similar apertures are needed 
to settle definitively this problem. 

\setcounter{table}{12}
\begin{table*}
\begin{minipage}{150mm}
\centering \caption{Comparison of optical and IR abundances for NGC~3603$^{\rm a}$}
\label{comparison}
\begin{tabular}{cccc}
\noalign{\hrule} \noalign{\vskip3pt}
Element/Ion & Opt CELs$^{\rm b}$ &  Opt CELs$^{\rm b}$ & IR CELs$^{\rm c}$ \\
& ({\ts} = 0.00) & ({\ts} $>$ 0.00) & \\
\noalign{\vskip3pt} \noalign{\hrule} \noalign{\vskip3pt}
O        & 8.46 $\pm$ 0.05  & 8.72 $\pm$ 0.05 & 8.79 $\pm$ 0.09 \\
N        & 7.62 $\pm$ 0.13  & 7.89 $\pm$ 0.14 & 7.96 $\pm$ 0.09 \\
S        & 7.03 $\pm$ 0.05  & 7.36 $\pm$ 0.08 & 7.12 $\pm$ 0.09 \\
Ne       & 7.76 $\pm$ 0.08  & 8.03 $\pm$ 0.11 & 8.08 \\
\noalign{\vskip3pt} \noalign{\hrule} \noalign{\vskip3pt}
O$^{++}$ & 8.42 $\pm$ 0.05  & 8.68 $\pm$ 0.08 & 8.61 $\pm$ 0.09 \\
S$^{++}$ & 6.83 $\pm$ 0.04  & 7.11 $\pm$ 0.09 & 6.90 $\pm$ 0.12 \\
Ne$^{++}$& 7.72 $\pm$ 0.05  & 8.00 $\pm$ 0.08 & 7.85 $\pm$ 0.09 \\
\noalign{\vskip3pt} \noalign{\hrule} \noalign{\vskip3pt}
\end{tabular}
\begin{description}
\item[$^{\rm a}$] In logarithmic units.
\item[$^{\rm b}$] This work.
\item[$^{\rm c}$] \citet{simpsonetal95}.
\end{description}
\end{minipage}
\end{table*}

\section{Conclusions}\label{conclu}

We present deep echelle spectroscopy in the 3100-10400 ${\rm \AA}$ range of bright zones of the Galactic 
{\hii} regions M16, M20 and NGC~3603. We have measured the intensity of about 250 lines per 
object. This is the most complete set of emission lines ever obtained for these three objects.

We have derived the physical conditions of each nebula making use of several line intensities and 
continuum ratios. The chemical abundances have been derived using the intensity of collisionally excited lines 
(CELs) for a large number of ions 
of different elements. We have determined, for the first time in the three objects, the C$^{++}$ and O$^{++}$ 
abundances from recombination lines (RLs) and, finally we have also determined the abundance of O$^{+}$ from RLs 
for the first time 
in M20.

We have obtained consistent estimations of the temperature fluctuations parameter, {\ts}, applying different methods: 
by comparing the O$^{+}$ (when available) 
and O$^{++}$ ionic abundances derived from RLs to those derived from CELs; by applying a chi-squared method 
which minimizes the dispersion of He$^+$/H$^+$ ratios from individual lines; and by comparing the electron 
temperatures derived from CELs to those derived from Balmer and Paschen continua. The adopted average value of 
{\ts} has been used to correct the ionic abundances derived from CELs.

We report the detection of several deuterium Balmer lines in the spectra of M16 and M20. The properties of these 
lines indicate that fluorescence is their most  probable excitation mechanism.

We have compared the results obtained for optical CELs in NGC~3603 with those obtained from far-infrared fine-structure 
CELs finding an apparent agreement, in spite of the high uncertainties of the abundances derived 
from far IR data,
if the temperature fluctuations paradigm is assumed. However, IR and optical 
spectrophotometry covering the same volume of the nebula is necessary to make a reliable and conclusive 
comparison between optical and IR CEL abundances.

\section*{Acknowledgments}

This work is based on observations collected at the European Southern Observatory, Chile, proposal number ESO 68.C-0149(A). 
We want to thank an annonymous referee for his/her comments, that have increased the quality of this work. 
JGR and CE would like to thank the members of the Instituto de Astronom\'{\i}a, UNAM, 
and of the INAOE, Puebla for their always warm hospitality. 
JGR would like to thank S. Sim\'on-D\'{\i}az and A. R. L\'opez-S\'anchez for 
fruitful discussions. This work has been partially funded by the Spanish 
Ministerio de Ciencia y Tecnolog\'{\i}a (MCyT) under projects AYA2001-0436 and AYA2004-07466. MP received partial 
support from DGAPA UNAM (grant IN
114601). MTR received partial support from FONDAP(15010003) and
Fondecyt(1010404). MR acknowledges support from Mexican CONACYT project J37680-E.


\end{document}